\documentclass{article}

\usepackage[english]{babel}
\usepackage[utf8]{inputenc}
\usepackage{amsmath}
\usepackage{graphicx}
\usepackage{epstopdf}
\usepackage[colorinlistoftodos]{todonotes}
\usepackage{authblk}
\usepackage{subcaption}
\usepackage[tiny]{titlesec}
\usepackage[margin=1.25in]{geometry}
\usepackage{cite}

\newcommand*\diff{\mathop{}\!\mathrm{d}}

\begin{document}

\title{\vspace{-1.2in}Recommendations and illustrations for the evaluation of photonic random number generators}

\author[1,2]{Joseph D. Hart}
\affil[1]{Institute for Research in Electronics and Applied Physics, University of Maryland, College Park, MD 20742}
 \affil[2]{Department of Physics, University of Maryland, College Park, MD 20742}
\author[3]{Yuta Terashima}
\affil[3]{Department of Information and Computer Sciences, Saitama University, Saitama, Japan}
\author[3]{Atsushi Uchida}
\author[4]{Gerald B. Baumgartner}
\affil[4]{Laboratory for Telecommunication Sciences, College Park, MD 20740}
\author[1,5]{Thomas E. Murphy}
\affil[5]{Department of Electrical and Computer Engineering, University of Maryland, College Park, MD 20742}
\author[1,2,6]{Rajarshi Roy}
 \affil[6]{Institute for Physical Science and Technology, University of Maryland, College Park, MD 20742}

\date{\today}

\maketitle

\begin{abstract}
The never-ending quest to improve the security of digital information combined with recent improvements in hardware technology has caused the field of random number generation to undergo a fundamental shift from relying solely on pseudo-random algorithms to employing optical entropy sources. Despite these significant advances on the hardware side, commonly used statistical measures and evaluation practices remain ill-suited to understand or quantify the optical entropy that underlies physical random number generation. We review the state of the art in the evaluation of optical random number generation and recommend a new paradigm: quantifying entropy generation and understanding the physical limits of the optical sources of randomness. In order to do this, we advocate for the separation of the physical entropy source from deterministic post-processing in the evaluation of random number generators and for the explicit consideration of the impact of the measurement and digitization process on the rate of entropy production. We present the Cohen-Procaccia estimate of the entropy rate $h(\epsilon,\tau)$ as one way to do this. In order to provide an illustration of our recommendations, we apply the Cohen-Procaccia estimate as well as the entropy estimates from the new NIST draft standards for physical random number generators to evaluate and compare three common optical entropy sources: single photon time-of-arrival detection, chaotic lasers, and amplified spontaneous emission.

\end{abstract}

Random number generation underlies modern cryptographic techniques used to ensure the privacy of digital communication and storage. In order to improve security, digital information systems have begun to utilize optical or other physical sources to generate high-speed unpredictable signals. However, the methods most commonly used to evaluate random number generators (RNGs) have not yet evolved to reflect the increasing importance of physical entropy sources for modern cryptography. 

Historically, random number generation has been dominated by algorithms that, given a seed, produce a sequence of pseudo-random numbers. However, since pseudo-random number generators (PRNGs) are deterministic algorithms, if an attacker is able to determine the seed, all security is lost. In order to defend against such problems, RNG designers are increasingly turning to physical means to either frequently re-seed or completely replace PRNGs, as evidenced by the commercialization of optical RNGs by companies such as PicoQuant \cite{picoquant}, IDQuantique \cite{IDQuantique}, and Whitewood Encryption \cite{hughes2016strengthening}.

Unlike PRNGs, physical processes can generate true randomness. Classical stochastic processes such as thermal or electrical noise can be used for entropy generation \cite{hamburg2012analysis}. Additionally, boolean chaos \cite{rosin2013ultrafast} and timing jitter in ring oscillators \cite{dichtl2007high,wold2009analysis} have been used to create electronic entropy sources. However, optical systems are especially well-suited for random number generation due resistance to external interference, speed, and access to quantum mechanical processes. Therefore, even though our analysis and recommendations are relevant for all physical entropy sources, we focus specifically on optical systems. 

The fundamental randomness of quantum mechanics present in many optical systems can be employed to generate true random numbers. In some optical entropy sources such as single photon measurements \cite{wayne2009photon,wahl2011ultrafast,IDQuantique,yan2014multi,yan2015high}, optical parametric oscillators \cite{marandi2012all}, and spontaneous Raman scattering \cite{collins2015random}, the measurements themselves are quantized. In others, such as those based on amplified spontaneous emission \cite{williams2010fast,li2011scalable,argyris2012sub,wei2012high,yamazaki2013performance,li2014random,zhang2016fast,hughes2016strengthening}, laser phase noise \cite{guo2010truly,qi2010high,xu2012ultrafast,yuan2014robust,nie2015generation,liu2016117}, quantum vacuum fluctuations \cite{gabriel2010generator, shen2010practical, symul2011real,jofre2011true,shi2016random}, and stimulated Raman scattering \cite{bustard2013quantum,england2014efficient}, an unpredictable analog waveform with quantum mechanical origins is sampled and digitized. In this Perspective, we will provide an in-depth analysis of one of each type: single photon time-of-arrival measurements and amplified spontaneous emission. Optical RNGs based on photon detection \cite{picoquant,IDQuantique} and spontaneous emission \cite{hughes2016strengthening} are now commercially available. See ref. \cite{herrero2016quantum} for a review of stochastic RNGs based on these and other optical systems. 

Chaotic systems amplify uncertainties in initial conditions and sources of intrinsic noise \cite{fox1991amplification,bracikowski1992amplification}; only in the last decade has this inherent unpredictability been harnessed for random number generation in the form of chaotic lasers \cite{uchida2008fast,kanter2010optical,oliver2013fast,yamazaki2013performance,virte2014physical,sakuraba2015tb,tang2015tbits,butler2016optical,wang2017minimal,ugajin2017real}. For a review of chaotic lasers including their applications to RNGs, see ref. \cite{sciamanna2015physics} and \cite{uchida2012optical}.  While the authors know of no commercially available physical RNGs based on chaotic lasers, new developments in photonic integrated circuits \cite{ugajin2017real} and real-time, high-speed bit streaming \cite{shinohara2017chaotic} for chaotic laser RNGs lay the groundwork for commercialization.

Physical sources of randomness and PRNGs are best used in complementary roles. Physical sources can provide true randomness, but the raw output of a physical source is typically biased and not uniformly distributed. PRNGs, on the other hand, can provide a binary sequence that is unbiased and uniformly distributed but completely deterministic. The most secure RNGs combine the benefits of both methods by using physical sources to seed PRNGs or other post-processing algorithms; such implementations are used by the Intel Secure Key (the RDRAND command), available in Ivy Bridge processors \cite{hamburg2012analysis}, and in the commercially available optical RNGs provided by PicoQuant \cite{picoquant}, IDQuantique \cite{IDQuantique}, and Whitewood Encryption \cite{hughes2016strengthening}. Official guidelines for how to combine a physical entropy source and a PRNG are currently under development by the U.S. National Institute of Standards and Technology (NIST) \cite{90C}.

Due to the increasing importance of the security of digital information and the wide variety of physical methods used to generate random numbers, a standard set of evaluation metrics for random number generation is essential. Previous works have used a variety of methods to estimate the entropy of physical sources of randomness \cite{xu2012ultrafast,wayne2009photon,wahl2011ultrafast,yan2015high,yan2014multi,nie2015generation,liu2016117,virte2014physical,sakuraba2015tb,tang2015tbits,wang2017minimal,ugajin2017real,li2014random,corron2016entropy,gabriel2010generator}; however, many of these techniques assume that there are no inter-sample correlations. As of this writing, there is no widely accepted technique to estimate the entropy of physical entropy sources. It is important that evaluation metrics and standards reflect the fundamental differences between PRNGs and physical entropy sources; however, common testing practices do not currently distinguish between the two. We are not the first to recognize these problems; indeed, NIST is currently developing a new set of standards and evaluation techniques specifically for physical entropy sources \cite{90B}. These new standards recommend entropy rate as the figure of merit for physical entropy sources.  

In this Perspectives article, we review the current practices in the evaluation of physical RNGs and call for a renewed emphasis on understanding the origin of and physical and information theoretical limitations on the randomness in the design, testing, and validation of optical entropy sources. We advocate for the separation of the physical entropy source from deterministic post-processing in the evaluation process and for the use of the $h(\epsilon,\tau)$ entropy rate analysis \cite{gaspard1993noise,hagerstrom2015harvesting}. The $h(\epsilon,\tau)$ entropy rate analysis emphasizes that the entropy rate is a function of measurement resolution $\epsilon$ and sampling period $\tau$. While the existing statistical tests used for physical RNG evaluation offer a simple ``pass or fail'' evaluation, the $h(\epsilon,\tau)$ analysis provides insight that is more relevant for the design of optical RNGs, including information about the physical origins of randomness and the impact of the digitization process on entropy extraction. Finally, we use the $h(\epsilon,\tau)$ analysis to compare three state-of-the-art optical entropy sources: single photon counting, chaotic lasers, and amplified spontaneous emission noise.

In section \ref{sec:evaluation}, we review current physical RNG evaluation practices and present our recommendations. In particular, section \ref{sec:currenteval} describes the current standards for physical RNG evaluation and some of its shortcomings. In section \ref{sec:NIST} we briefly describe the new NIST draft recommendations for the evaluation of physical random number generators, and we provide our own recommendations in section \ref{sec:dynamicalsystems}. In section \ref{sec:opticalsourcesreview}, we review three different methods of optical entropy generation and present the results of our own measurement and evaluation of these entropy sources. We provide some concluding thoughts and an outlook to the future of physical RNG in section \ref{sec:conclusion}. 

\section{\label{sec:evaluation}Evaluation of Physical Random Number Generators}
Evaluating a PRNG is relatively straightforward: NIST has published specific guidelines for the design and testing of PRNGs \cite{90A}. In contrast, physical RNGs are much more difficult to evaluate due in part to the wide variety of physical processes that can be used \cite{90B}. 

\subsection{\label{sec:currenteval}State of the art}
\begin{figure*}
\centering
\includegraphics[width=0.85\textwidth]{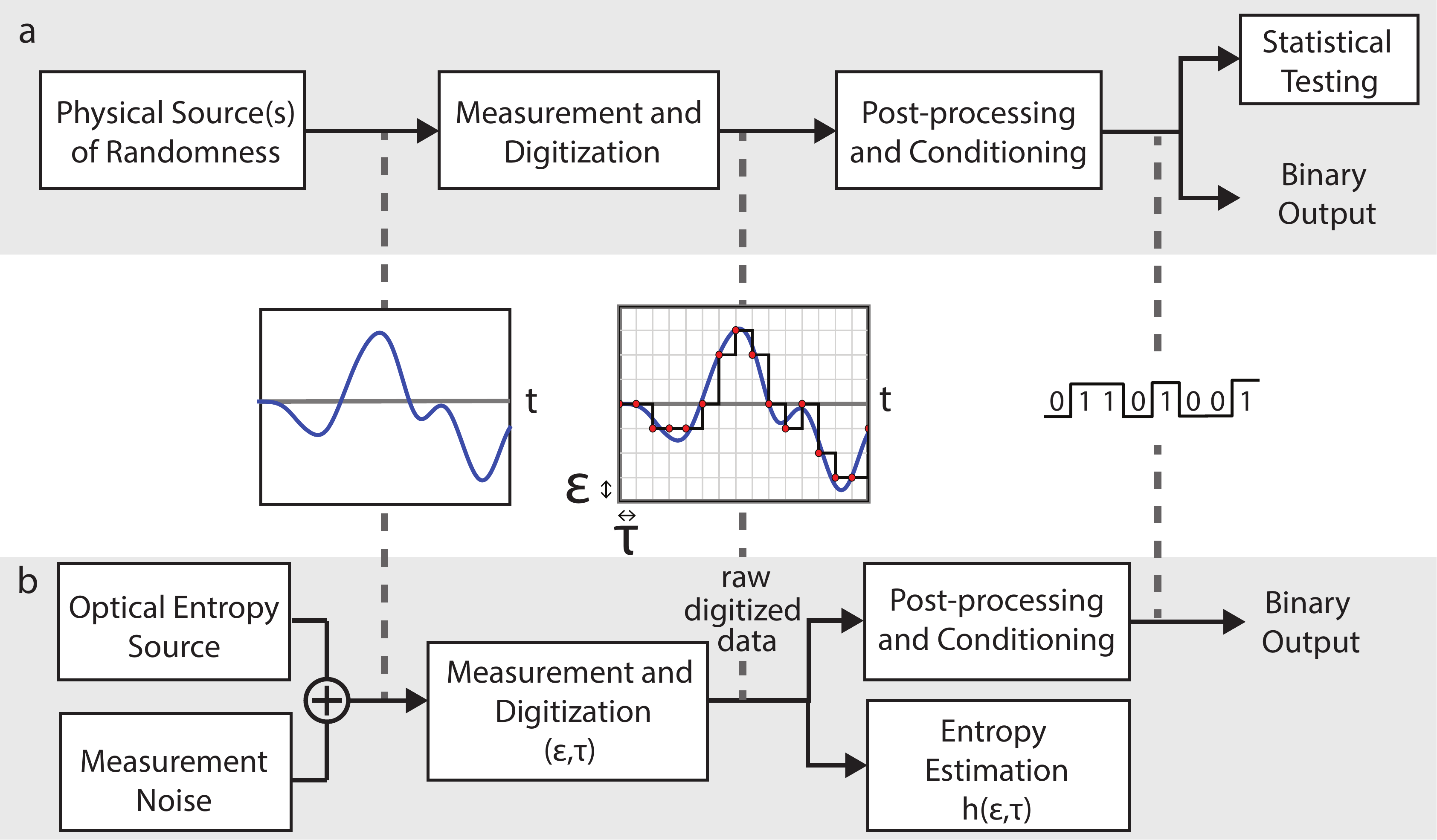}

\caption{\label{fig1}Methods of optical RNG testing and evaluation. a) Current practice involves performing statistical tests on the final, post-processed output bit sequence. b) Our recommendation is to use the raw digitized data to make an estimate of the rate at which one is entitled to harvest entropy, then use appropriate post-processing to extract that entropy from the digitized data. The measurement parameters $\epsilon$ and $\tau$ and the post-processing method should be carefully chosen such that entropy is extracted from the desired physical entropy source rather than from measurement noise. The sample time series show an analog signal, a digitized signal with measurement parameters $\epsilon$ and $\tau$, and a post-processed output bit sequence. }
\end{figure*}
The most common procedure used to evaluate physical RNGs, depicted in Fig. 1a, is the following: collect data from the physical system; perform deterministic post-processing (such as von Neumann's method) and/or conditioning (such as least significant bit extraction \cite{oliver2013fast}, exclusive or (XOR) \cite{uchida2008fast}, and hashing \cite{wayne2009photon}) on the data in order to remove bias and whiten the output; and run a suite of statistical tests (such as the NIST suite \cite{oldNIST} or DIEHARD \cite{diehard}) on the output bit sequence. The distinction between post-processing and conditioning is defined somewhat arbitrarily by NIST \cite{90B}, as discussed in the next section. The rate of random bit generation claimed is typically the highest possible rate such that the output bit sequence can pass the suite of statistical tests. As we will discuss, this method of evaluating physical RNGs has significant shortcomings.

\begin{figure*}
\includegraphics[width=0.95\textwidth]{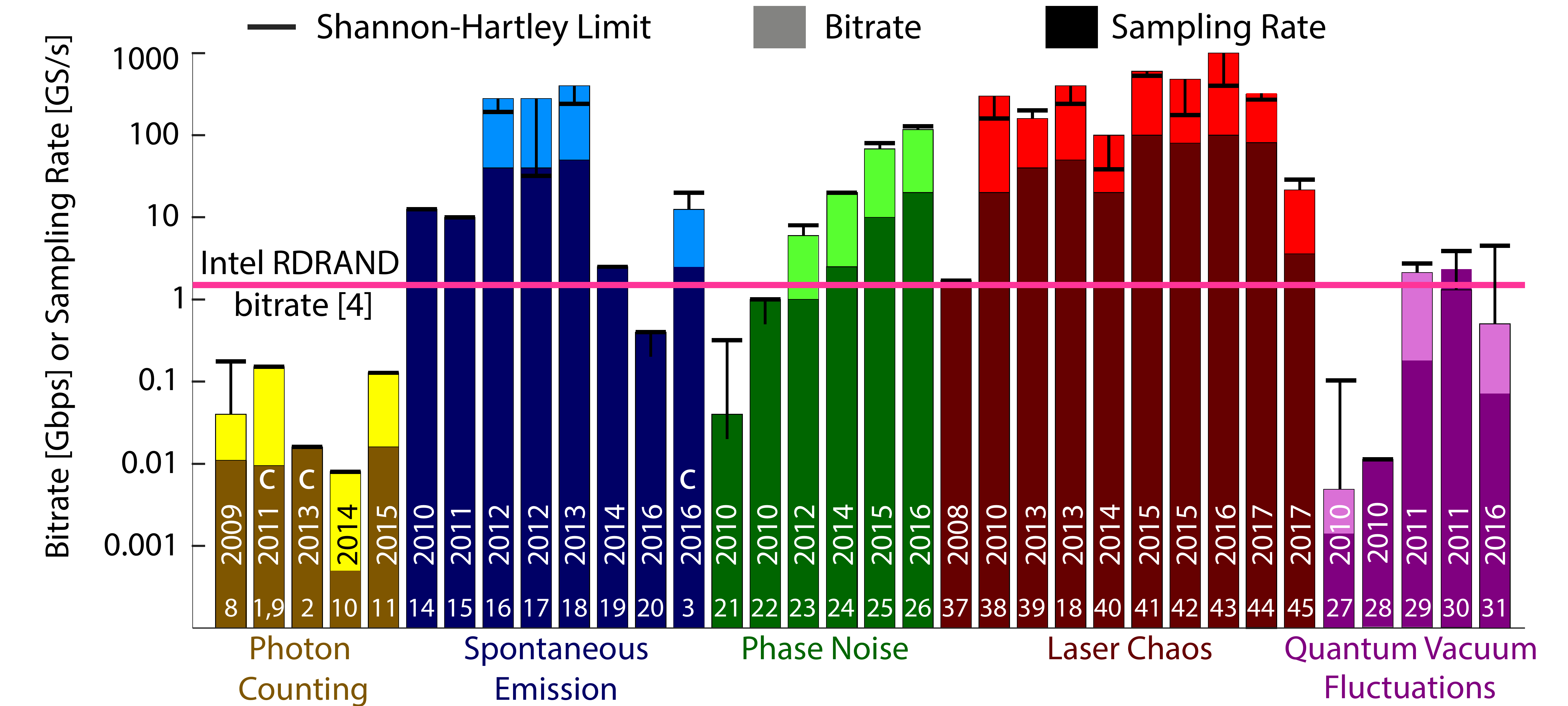}
\caption{Survey of recent optical random number generation rates. The darker vertical bars indicate the sampling rate used and the lighter bars indicate the claimed random bit rate. The number of the reference is written on each bar. The black bars indicate the Shannon-Hartley limit given by Eq. \ref{eq:shannonhartley}. Commercial products are denoted by a C. }
\end{figure*}

Figure 2 provides a survey of some recently published results using five common optical techniques for random number generation. The bit rates given are those claimed by the authors, and the sampling rate is the number of samples per second measured by the digitizer. In many cases the claimed bit rate is higher than the sampling rate; this is a result of the digitizer obtaining more than 1 bit per sample (e.g., by using an 8-bit analog to digital converter).

An information theoretical upper bound on the entropy rate is given by the Shannon-Hartley limit \cite{desurvire2009classical}
\begin{equation}
\label{eq:shannonhartley}
h_{SH}=2 \textup{BW}\cdot N_{\rm \epsilon},
\end{equation}
where $N_{\rm \epsilon}$ is the number of bits per sample that the digitizer measures at a given measurement resolution $\epsilon$, and $BW$ is the bandwidth of the signal measured by the digitizer. $BW$ is limited by the analog bandwidth of the physical entropy source as well as the detectors and digitizer. Eq. \ref{eq:shannonhartley} gives the maximum rate at which information can be obtained from the signal by the digitizer \cite{desurvire2009classical}. Of course, Eq. \ref{eq:shannonhartley} overestimates the upper bound because the effective bandwidth is often less than the standard signal bandwidth \cite{lin2012effective} and the effective number of bits of a digitizer is often less than the stated number of bits \cite{walden1999analog}.

Fig. 2 reveals a significant shortcoming in the current practice of quantifying optical random number generation: several of the recently reported RNG systems, while producing data that passes all of the existing statistical tests, achieve a rate that exceeds even our overestimate of the Shannon-Hartley limit (horizontal black bars).  In most of the violating cases, the sampling rate is higher than the Nyquist rate (2BW) of the signal, resulting in strong inter-sample correlations. Post-processing may obscure these correlations from statistical tests; however, because post-processing is deterministic it cannot increase the rate of entropy production.

When one considers that many PRNGs pass the statistical tests, it is not surprising that those statistical tests can be passed by post-processing the output of a physical system, even if that output is not random. While statistical tests can provide some assurance of statistical uniformity, they provide no guarantee that there is no underlying pattern that could later be discovered. Therefore, statistical tests are perhaps best viewed as a sanity check against blatant errors, rather than a proof of randomness.

Statistical test suites are also limited to a simple ``pass or fail'' evaluation that provides little insight into the physical processes generating the random numbers. Choosing a physical process that can be theoretically justified as random and then showing that the measured entropy is actually coming from that random process provides much greater assurance of unpredictability than can simply putting a sequence of bits through statistical tests that any good pseudo-random number generator will pass.

While there is nothing wrong with having the sole aim of passing statistical tests, physical RNGs that do this while violating information theory limits are perhaps better called physical-based PRNGs, rather than true physical RNGs, as suggested in refs. \cite{corron2016entropy} and \cite{li2014two}. 

The upper bound on entropy harvesting provided by Eq. \ref{eq:shannonhartley} assumes that the probability density function (PDF) is uniform, the maximal entropy distribution. For small $\epsilon$ (high resolution), one can find a more stringent upper bound on the entropy rate by accounting for the fact that PDFs found in physical systems are generally not uniform \cite{gaspard1993noise}:

\begin{figure}
\includegraphics[width=0.95\textwidth]{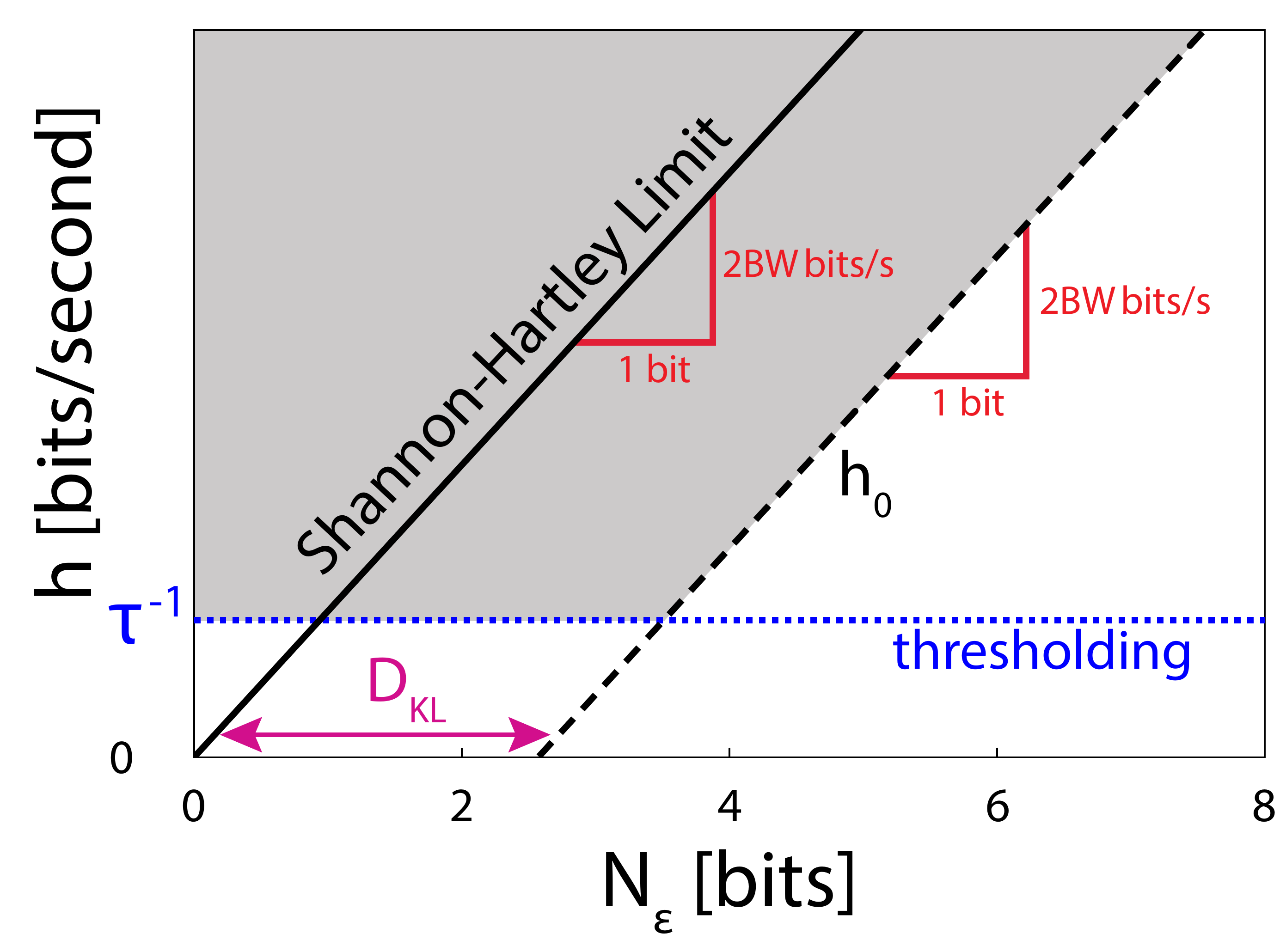}
\caption{Illustration of information theoretical limits for physical RNG. The Shannon-Hartley limit (solid black line, Eq. \ref{eq:shannonhartley}) is the theoretical upper limit for rate of information transfer for a system with a given bandwidth. $h_0$ (dashed black line, Eq. \ref{eq:h0}) is a correction to the Shannon-Hartley limit obtained by accounting for non-uniformity of the PDF of the physical process generating the entropy. The Shannon-Hartley limit has a slope of $\tau_{max}^{-1}=2BW$ bits per second per bit of resolution, while the slope of $h_0$ is $\min(\tau^{-1},2BW)$ bits per second per bit of resolution. Most RNG designers want to push the limit of random bit generation, so here we show the slope of $h_0$ as the maximum of 2$BW$ bits per second per bit of resolution. The x-intercept of $h_0$ is given by the Kullback-Leibler divergence of the uniform distribution from the experimental probability distribution. The $h_0$ limit is only valid for fine resolution; for IID systems an entropy rate of $\tau^{-1}$ bits per second can always be obtained by thresholding ($N_{\rm\epsilon}=1$, blue dotted line). The sampling rate for thresholding is also limited by the Nyquist rate $2BW$. The region of entropy rates that is unobtainable according to the limits provided by Eqs. \ref{eq:shannonhartley} and \ref{eq:h0} is indicated by the gray shading.}
\label{fig:limitfig}
\end{figure}

\begin{equation}
\label{eq:h0}
h_0 = \min(\tau^{-1}, 2BW)\big(N_{\rm \epsilon}-D_{KL}(p(x)||u(x))\big), 
\end{equation}
where $p(x)$ is the PDF of the physical entropy source, $u(x)$ is the uniform distribution over the interval of $x$-values for which $p(x)$ is non-zero, and $D_{KL}(p(x)||u(x))$ denotes the relative entropy or Kullback-Leibler divergence \cite{coverthomas} of $u(x)$ from $p(x)$, and  $\tau^{-1}$ is the sampling rate; according to the Nyquist theorem, one cannot obtain more entropy by sampling faster than 2$BW$ \cite{desurvire2009classical}. Eq. \ref{eq:h0} simply says that the maximum obtainable entropy rate is the maximum sampling rate times the average entropy per sample. For further discussion on the $h_0$ limit, see the Appendix.

We emphasize that Eq. \ref{eq:h0} is valid only for fine measurement resolution; in the case of thresholding ($N_{\rm \epsilon}$=1 bit), one can always obtain $h(N_{\rm \epsilon})=\tau^{-1}$ bits/s for any independent, identically distributed (IID) random process by setting the threshold at the median. The sampling rate for thresholding is also limited by the Nyquist rate $2BW$.

Figure \ref{fig:limitfig} illustrates the relationships between these information theoretical limitations on obtainable entropy rates. The limits in Eq. \ref{eq:h0} are information theoretical limits that depend on the specifications of the measurement apparatus and on the bandwidth and PDF of the physical system; an additional physical limit, the Kolmogorov-Sinai (metric) entropy rate, exists for dynamical (chaotic) systems. We discuss the Kolmogorov-Sinai entropy rate in sections \ref{sec:hepstau} and \ref{sec:laserchaos}.

\subsection{\label{sec:NIST}New NIST Draft Recommendations}
The NIST Draft Recommendation for the Entropy Sources Used for Random Bit Generation \cite{90B} tries to resolve some of these problems by separating out the algorithmic, pseudo-random parts of random number generation from the physical entropy source. It also gives recommendations on how to combine the pseudo-random algorithm and the entropy source once they have been separately validated \cite{90C}. The NIST draft recommendation also requires a justification of how the entropy source works and why it produces acceptable entropy.

 The NIST draft standards are based on an entropy source model similar to the one shown in Fig. 1b; The only difference is that it does allow for some simple post-processing techniques to be applied to the raw digitized data before estimation of the entropy rate. NIST distinguishes post-processing (only von Neumann's method, linear filtering method, or length-of-runs method) from conditioning (such as some hash functions), which has fewer restrictions and is not allowed to increase the entropy estimate.

For entropy sources that are potentially not IID, the entropy estimation procedure is quite simple. Run two suites of tests on the (post-processed but not conditioned) data. The first suite of tests estimates the min-entropy \cite{konig2009operational} per sample in the data set; the min-entropy is designed to provide a conservative estimate of the entropy. The second suite of tests is a pass/fail set of so-called ``Restart Tests'': These ensure that the entropy source does not behave the same way each time it is restarted. We will discuss only the entropy estimation suite. If a conditioning procedure is used, one should adjust down the min-entropy estimate if appropriate, as described in the NIST recommendation \cite{90B}. The resulting estimate of min-entropy per sample gives the upper limit on the rate at which entropy can be extracted from the source.

The min-entropy estimation suite includes 10 different tests, and the final min-entropy per sample estimate is the minimum of all the estimates. We have run all 10 tests on our data, but since most of them give similar results we discuss only two of the estimates here: the most common value (MCV) estimate and the Markov estimate. The simplest entropy estimate is the MCV estimate. It assumes the signal has no correlations and estimates the entropy as $-\log_2(p_{max})$, where $p_{max}$ is the fraction of samples appearing in the most common bin. The second entropy estimate is the Markov estimate, which takes into account first-order correlations. For a complete description of all the tests, see ref. \cite{90B}.

We support NIST's efforts to separate the evaluation of pseudo-random algorithms and physical entropy sources. The requirement to justify the workings of the entropy source and to provide an explicit statement of the expected entropy rate places long overdue emphasis on understanding the physics of the entropy source. 

The NIST draft recommendation allows some post-processing before the entropy analysis, but deterministic algorithms cannot increase the entropy rate and serve only to make the entropy estimation process more difficult. Indeed, this was recently shown for the commonly used post-processing technique of least significant bit extraction \cite{corron2016entropy}. It is known that some of the tests in the NIST suite severely underestimate the min-entropy of entropy sources with normal distributions \cite{kelsey2015predictive}; this is a real problem because many of the best physical entropy sources have normal or approximately normal PDFs. Thus, the NIST test suite unintentionally encourages designers of these systems to include post-processing before testing, since this is the only way their source can receive a high entropy estimate from the NIST suite. The NIST draft recommendation does not address the details of the digitization process, which has a bandwidth due to the detection apparatus, measurement resolution $\epsilon$, and sampling frequency $\tau^{-1}$ that all impact the rate at which entropy can be harvested from a physical system. Finally, the NIST draft standards do not mention chaotic entropy sources, even though it is well-known that entropy can be harvested from chaotic systems due to their sensitive dependence on initial conditions \cite{sciamanna2015physics,uchida2008fast,boffetta2002predictability}.
In light of these concerns, we provide some additional recommendations to designers and evaluators of physical entropy sources in the next section.

\subsection{\label{sec:dynamicalsystems}Recommendation: A dynamical systems approach to entropy estimation}
Both stochastic and chaotic physical systems have been used to generate entropy at high rates. It is therefore important to have techniques that can accurately estimate the entropy from both stochastic and chaotic sources. For this, we recommend a dynamical systems approach to entropy estimation.

Our first recommendation regards the role of post-processing in the evaluation of RNGs. As we have previously mentioned, deterministic post-processing algorithms are useful for extracting entropy but cannot increase the entropy production rate of the physical source. In order to obtain a more accurate and insightful measure of the suitability of a physical system for random number generation, we recommend estimating the rate of entropy production \textit{directly from the raw digitized data}. Post-processing and conditioning techniques can then be chosen to extract random bits from the physical source at a rate up to the entropy rate. This procedure is depicted in Fig. 1b.

As described above, many different physical processes can generate entropy. It is even possible for a single system to have multiple sources of entropy; for example, a chaotic laser entropy source might have entropy from the chaotic dynamics (which amplify intrinsic quantum mechanical noise from spontaneous emission) and from electronic noise in the detector. We therefore recommend that designers take care in choosing measurement parameters--measurement resolution $\epsilon$ and sampling period $\tau$--and post-processing techniques that extract entropy from the desired physical source.
In the rest of this section, we present $h(\epsilon,\tau)$ as a technique to guide these choices.

\subsubsection{\label{sec:hepstau}Noise, chaos, and $h(\epsilon,\tau)$}

Gaspard and Wang \cite{gaspard1993noise} have shown that $h(\epsilon,\tau)$ estimated by the Cohen-Procaccia algorithm \cite{cohen1985computing} can be used to directly compare the entropy production of stochastic and chaotic processes. $h(\epsilon,\tau)$ treats the entropy rate as a function of the measurement resolution $\epsilon$ and the sampling rate $\tau^{-1}$. Such an analysis can provide insight into the type of physical process (stochastic or chaotic) that is generating the entropy at a given measurement resolution or time scale. For example, in a deterministic chaotic system as the measurement resolution $N_{\rm \epsilon}$ increases, $h(\epsilon,\tau)$ approaches a constant given by the Kolmogorov-Sinai entropy rate; however, in a purely stochastic system, $h(\epsilon,\tau)$ scales as $N_{\rm \epsilon}$ \cite{gaspard1993noise}. These predictions were recently verified experimentally \cite{hagerstrom2015harvesting}. In the context of physical RNGs, the $h(\epsilon,\tau)$ analysis can guide the choice of the best $\epsilon$, $\tau$, and post-processing method to extract entropy from the desired physical source. 

We note that in all cases, the sampling rate $\tau$ is the embedding delay. For more details on the Cohen-Procaccia algorithm and its ability to accurately estimate the entropy rates for stochastic and chaotic processes, see the Appendix and ref. \cite{gaspard1993noise}. 

The $h(\epsilon,\tau)$ analysis is not limited to continuous time series; it can also be applied to physical entropy sources based on discrete events such as single photon detection. For example, entropy sources using single photon time-of-arrival measurements \cite{wayne2009photon,wahl2011ultrafast,yan2015high} can be analyzed by considering the entropy rate as a function of the temporal precision ($\epsilon$) of the measurement of the arrival times and of the maximum count rate ($\tau^{-1}$), as we show in the section \ref{sec:PC}

Despite these important advantages, the $h(\epsilon,\tau)$ analysis has only recently been applied in the context of physical random number generation \cite{hagerstrom2015harvesting}. The RNG system in ref. \cite{hagerstrom2015harvesting} was designed to study the ($\epsilon,\tau$) entropy rate estimate on low-dimensional chaotic experimental data and generated only a few hundreds of bits of entropy per second. In the following sections, we use the $h(\epsilon,\tau)$ analysis to compare three state-of-the-art optical RNG techniques: single photon time-of-arrival measurements, digitization of chaotic laser data, and digitization of amplified spontaneous emission noise data. An $h(\epsilon,\tau)$ analysis can be performed with any entropy estimation method; we use the Cohen-Procaccia estimate since it has been shown to work for both chaotic and stochastic sources of entropy \cite{gaspard1993noise}. While the NIST draft \cite{90B} does not discuss $h(\epsilon,\tau)$, we also compute an $h(\epsilon,\tau)$ using the MCV and Markov estimates from the NIST suite \cite{90B} in order to compare with the Cohen-Procaccia estimate.

\section{\label{sec:opticalsourcesreview}Review of some optical entropy sources}
\subsection{\label{sec:PC}RNG with Single Photon Detection}

The detection of single photons is perhaps the most established optical RNG technique. There are many different techniques for generating randomness from single photon detection; for a recent review, see ref. \cite{herrero2016quantum}. Perhaps the most straightforward way is to send a single photon through a 50:50 beam splitter and assign a ``0'' if it is detected at one output port and a ``1'' if it is detected at the other; this is the method used by the commercial RNG from \textit{ID Quantique} \cite{IDQuantique}, and provides 1 bit per photon of entropy. Another method is to count the number of photons $n$ detected in a given time window from a low intensity light source. If the average photon interarrival time is much less than the detector dead time, $n$ will be a random variable that follows the Poisson distribution \cite{herrero2016quantum}. It turns out that this technique has the same rate of entropy production as the one we focus on here: single photon time-of-arrival measurements \cite{wayne2009photon,wahl2011ultrafast,yan2015high}.

\begin{figure}
\centering
\includegraphics[width=.95\textwidth]{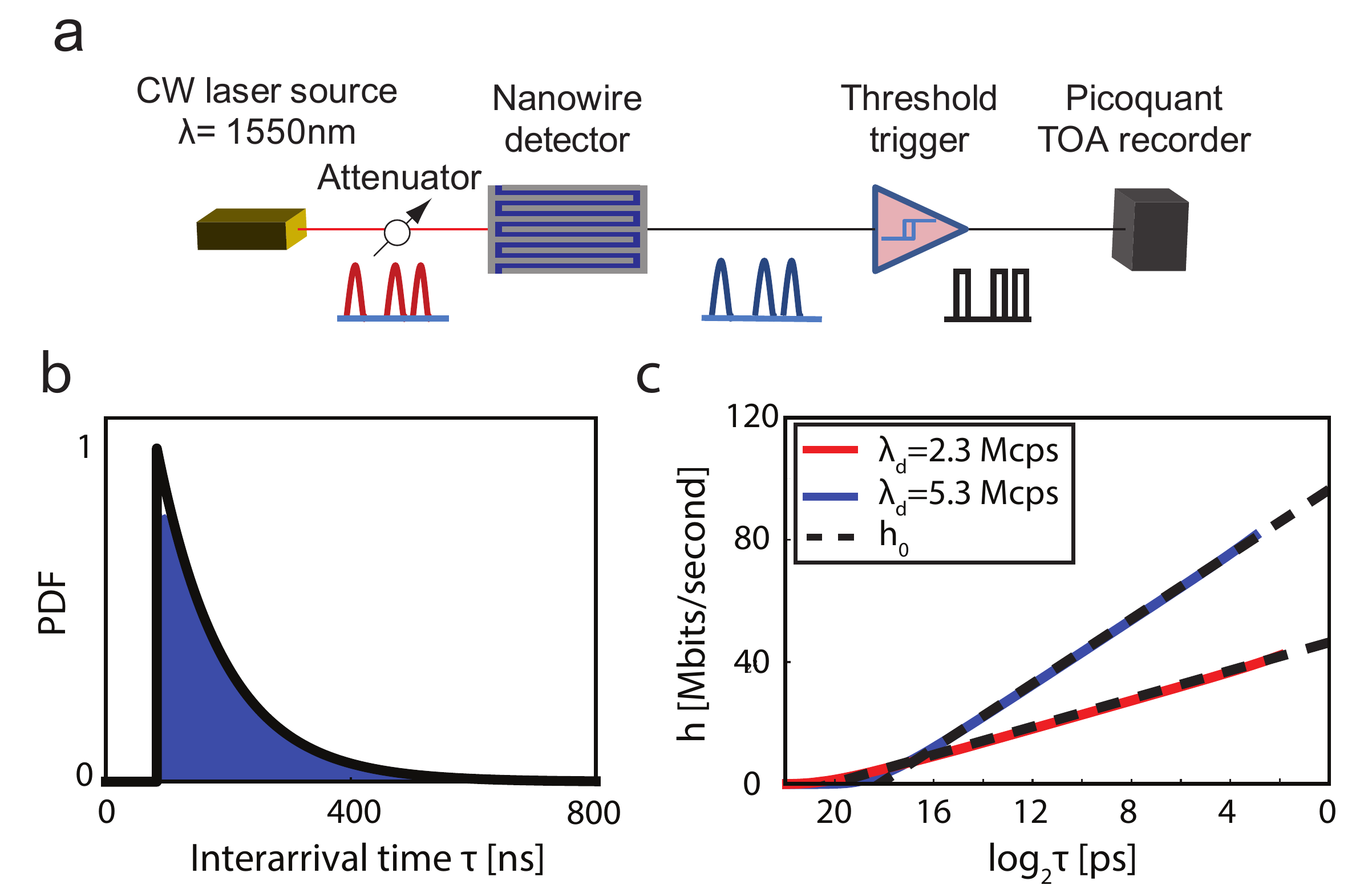}
\caption{a) Apparatus used to make single photon time-of-arrival measurements with a precision of 1 ps. b) Experimental histogram (shaded) and theoretical PDF (black line). The PDF is given by $p(\Delta t)=\lambda_d\exp{[-\lambda(\Delta t-\tau_d)]}$ where $\lambda_d=5.37$ Mcps and $\tau_d$=88 ns. c) Entropy rate $h$ as a function of timing precision for single photon time-of-arrival measurements. Two different detected photon rates $\lambda_d$ were used: 2.3 Mcps and 5.37 Mcps. Eq. \ref{eq:PCrate2} is used for $h_0$. In both cases, the Cohen-Procaccia entropy rate estimate agrees excellently with the prediction from Eq. \ref{eq:PCrate2}.}
\label{fig:PCapparatus}
\end{figure}

Fig. \ref{fig:PCapparatus}a depicts our experimental realization of high-precision time-tagged photon counting. The optical output of a 1550 nm CW laser is attenuated to an average photon rate of several million photons per second. Photons are detected using superconducting nanowire detectors, and single photon arrivals are time-tagged by the HydraHarp 400 (PicoQuant) with a precision of 1 ps. The digitized interarrival times between consecutive photons serve as our random signal.

The photon arrivals can be described as a Poisson process with constant rate $\lambda$: the probability per unit time for a photon to arrive is constant and independent of previous photon arrivals. It is well-known that the interarrival times of a constant rate Poisson process follow an IID exponential distribution of the form $p(t)=\lambda\exp{[-\lambda t]}$ \cite{herrero2016quantum,wayne2009photon,wahl2011ultrafast}. We can calculate the entropy per photon generated from these interarrival times as a function of the time-tagging resolution $\tau$ using Shannon's definition of entropy:
\begin{equation}
H(\tau)=-\sum_{k=0}^\infty p_k(\tau)\log_2(p_k(\tau)),
\end{equation}
where $p_k=\exp{[-k\lambda\tau]}(1-\exp{[-\lambda\tau]})$ is the probability of the photon interarrival time occurring between time $k\tau$ and $(k+1)\tau$. This can be evaluated in closed form as
\begin{equation}
\label{eq:PCentropy}
H(\tau)=\frac{(1-p_0)\log_2(1-p_0)}{p_0}+\log_2(p_0),
\end{equation}
where $p_0\equiv1-\exp{[-\lambda\tau]}$. The entropy generation rate $h=\lambda H$. 

Of course, real single photon detectors have a dead time. For non-paralyzable detectors, the dead time $\tau_d$ does not affect the shape of the PDF; it only shifts it by $\tau_d$, as shown in Fig. \ref{fig:PCapparatus}b. This does not affect $H$, the entropy per photon. The dead time does, however, affect the average rate of photons that are detected: $\lambda_d=\lambda/(1+\lambda\tau_d)$ \cite{wahl2011ultrafast}. Thus the entropy rate becomes $h=\lambda_dH$. Our detector \textit{is} paralyzable; however, if the photon rate is not too high, we can approximate it as non-paralyzable. Further, if the probability of more than one photon arriving in a single time bin is small ($\lambda\tau<<1$), we can approximate the entropy rate for photon time-of-arrival measurements as 
\begin{equation}
\label{eq:PCrate2}
h(\tau)=-\lambda_d\log_2(\frac{\lambda\tau}{e}).
\end{equation}
This is exactly the same entropy rate one would obtain by using Eq. \ref{eq:h0}, with $\min(\tau^{-1}, 2BW)=\lambda_d$.

We estimate the entropy rate of experimentally measured photon interarrival time measurements using the Cohen-Procaccia algorithm as a function of the time-tagging resolution $\tau$. We have performed time-of-arrival measurements for two different detected rates: $\lambda_d$=2.3 Mcps and $\lambda_d$=5.37 Mcps. In the first case the dead time is not important ($\lambda_d\approx\lambda$), while in the second case the dead time causes a loss of about 40\% of the photons.  In both cases these results give excellent agreement with Eq. \ref{eq:PCrate2}, as shown in Fig. \ref{fig:PCapparatus}c. Furthermore, only 1 dimension is needed for the Cohen-Procaccia algorithm to converge. This suggests that the photons were indeed generated by a Poisson process with no intersample correlations.

The entropy rates obtainable from modern photon counting experiments are on the order of a few hundred Mbits/s, and physical limitations present a significant challenge for improving performance. As shown in Eq. \ref{eq:PCrate2}, $h$ scales \textit{logarithmically} with the time-tagging resolution, suggesting that increasing the time-tagging resolution beyond the current state-of-the-art of 1ps is not an economical way to improve performance. The most efficient way to increase the entropy rate is to increase the detected photon rate $\lambda_d$, which is limited by the detector dead time and is typically on the order of a few tens of millions of counts per second with current technology. It is of course possible to increase the entropy rate by combining techniques. For example, Stipcevic and Bowers were able to obtain one additional bit per photon by combining the 50:50 beam splitter method with time-of-arrival measurements \cite{stipcevic2015spatio}. However, it is clear that this does not improve the scaling.

Single photon detection techniques are attractive because of their quantum mechanical nature and conceptual simplicity. However, in light of the physical limitations described above, we find it unlikely that RNG using single photon detection techniques can become competitive with the high-speed digitization of unpredictable analog waveforms, which can produce entropy rates of hundreds of Gbits/s.

\subsection{\label{sec:laserchaos}RNG from Chaotic Lasers}
It has long been known that a semiconductor laser can be made chaotic by creating a time-delayed optical feedback via a reflector  \cite{lang1980external,mork1992chaos}. In 2008 it was demonstrated for the first time that these chaotic lasers can be used to generate random numbers, and could do so at rates an order of magnitude faster than any previous physical RNGs \cite{uchida2008fast}. Since then, there has been much progress in building faster chaotic RNGs \cite{kanter2010optical,oliver2013fast,yamazaki2013performance,virte2014physical,sakuraba2015tb,tang2015tbits}.

Chaotic systems produce entropy by magnifying the small uncertainties in the initial conditions. The maximum rate at which entropy can be harvested from a chaotic system is called the Kolmogorov-Sinai entropy rate $h_{KS}$, which is equal to the sum of the positive Lyapunov exponents \cite{gaspard1993noise,boffetta2002predictability}. Ref. \cite{boffetta2002predictability} provides an in-depth review of the relationship between the Kolmogorov-Sinai entropy rate and the Shannon entropy rate.  Of course, real dynamical systems also have intrinsic noise sources, which fundamentally limit the precision with which initial conditions can be measured.  For the chaotic laser systems used for optical random number generation, spontaneous emission noise (which is of quantum mechanical origin) and potentially other noise sources are continuously amplified by the chaotic dynamics and contributes to the entropy production. The rate at which entropy can be harvested from a chaotic laser is limited both by the bandwidth of the detectors (Eq. 2) and $h_{KS}$ of the chaos.

\begin{figure}
\centering
\includegraphics[width=.95\textwidth]{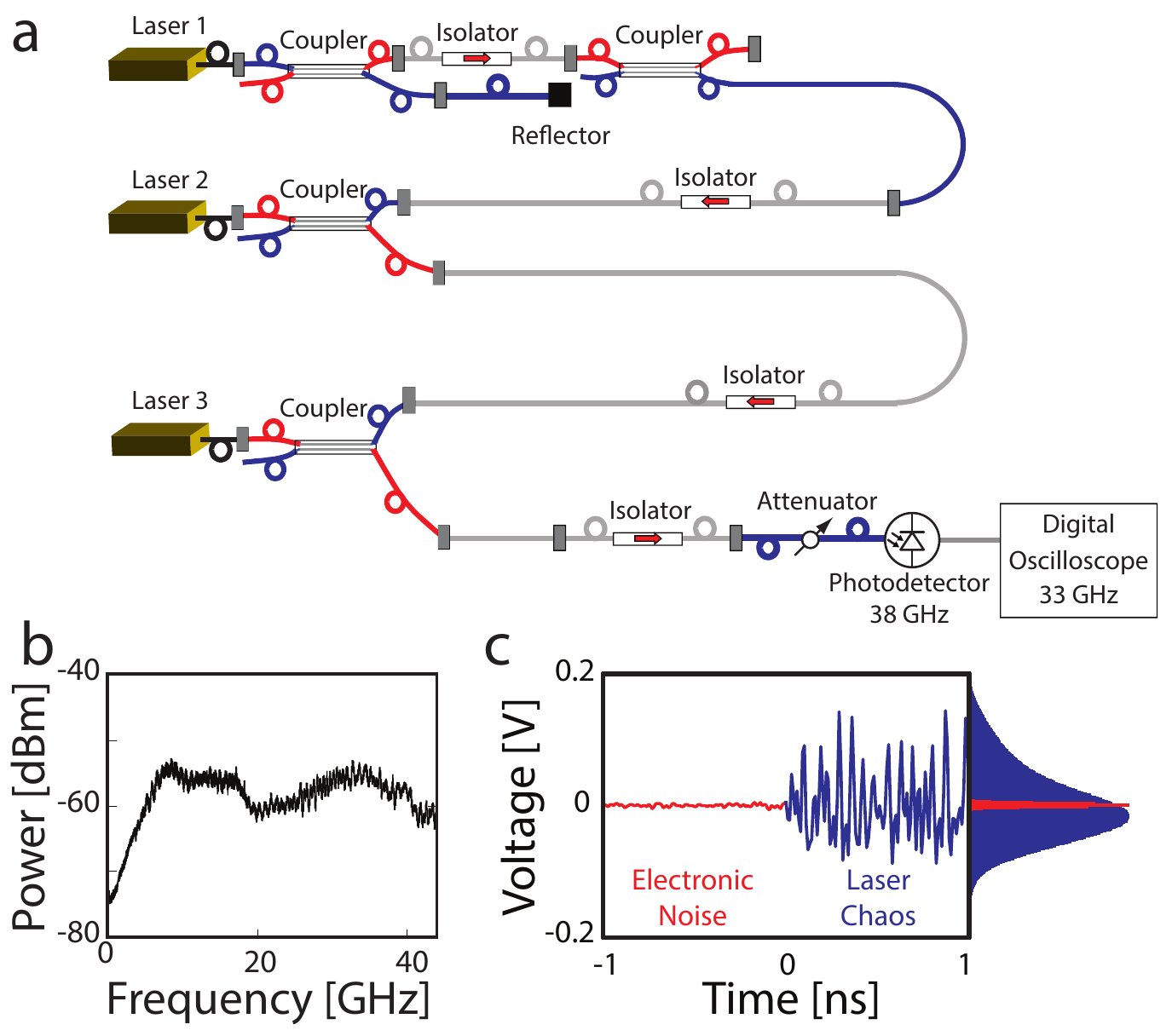}
\caption{a) Schematic of a three cascaded semiconductor laser entropy source. b) RF power spectrum of chaotic laser system. The injection currents of lasers 1, 2, and 3 are 58.5 mA, 59.0 mA, and 59.0 mA, respectively. The standard bandwidth is 34 GHz. c) Time series and PDF of the chaotic laser system (blue) and the electronic noise (red). The electronic noise is measured with all optics turned off.}
\label{fig:laserapparatus}
\end{figure}

The chaotic laser system we consider here obtains an enhanced bandwidth by cascading three semiconductor lasers (NTT Electronics, KELD1C5GAAA), as described in detail in ref. \cite{sakuraba2015tb} and shown in Fig. \ref{fig:laserapparatus}a. The first laser has time-delayed optical feedback from the reflector. The chaotic output intensity of the first laser is injected into the second laser, and the chaotic output of the second laser is then injected into the third laser. This cascading increases the standard bandwidth from 12.5 GHz at the first laser to 34 GHz at the final output. The final output intensity is detected by a 38 GHz photodetector (New Focus, 1474-A). The electrical signal from the photodetector is sampled by a high speed 8-bit oscilloscope (Tektronix DPO73304D, 33 GHz bandwidth, 100 GigaSamples/s).  The RF power spectrum is shown in Fig. \ref{fig:laserapparatus}b, and a typical time series is shown in Fig. \ref{fig:laserapparatus}c. One can immediately tell from the time series in Fig. \ref{fig:laserapparatus}c that a few of the least significant bits of the signal are due to electronic noise rather than the optical signal.

\begin{figure}
\centering
\includegraphics[width=0.7\textwidth]{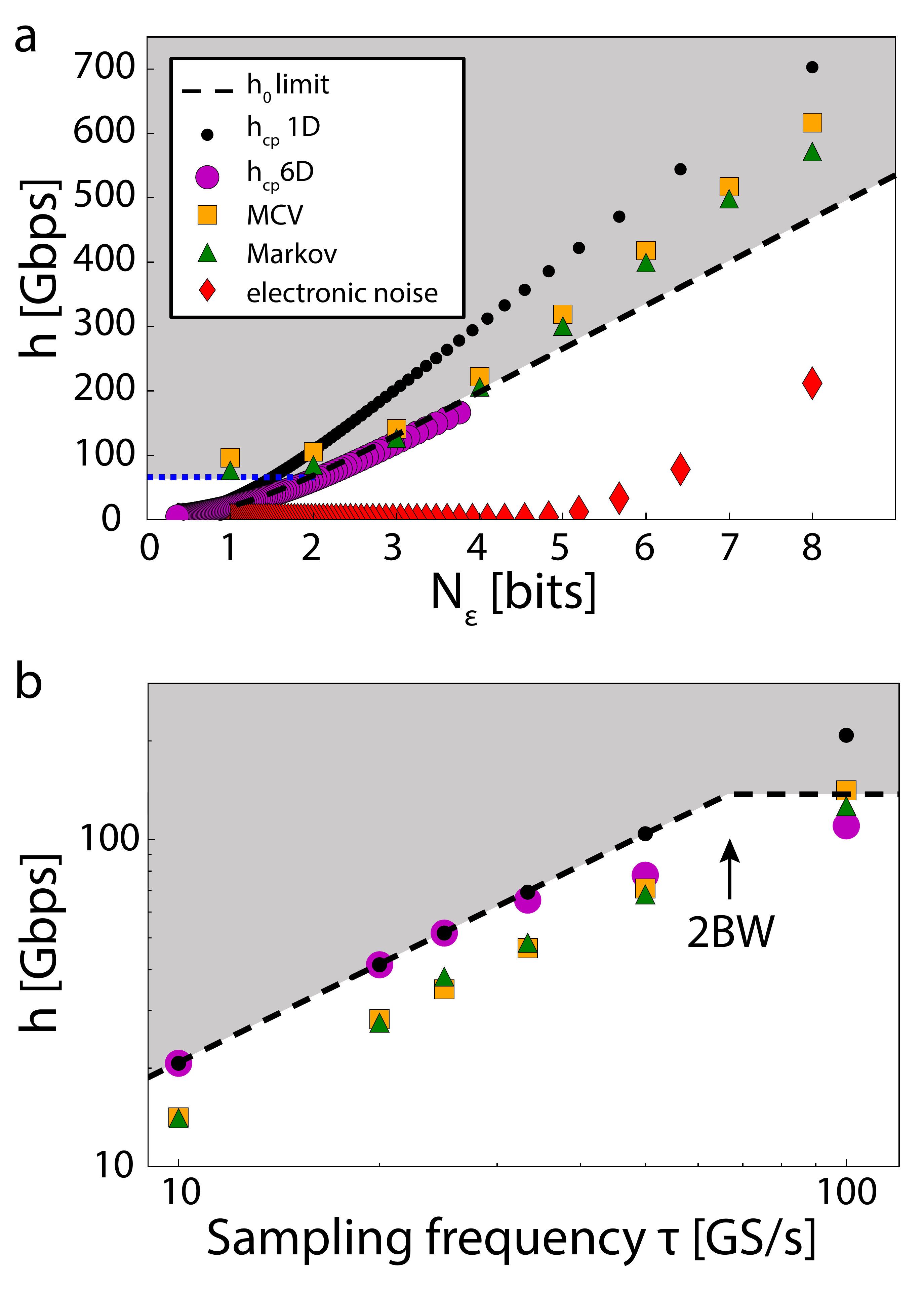}

\caption{a) Entropy rate $h(\epsilon)$ of the chaotic laser system for a fixed $\tau^{-1}=100$ GSamples/s. The blue dotted line is the entropy rate obtained by thresholding at the median, $\tau^{-1}$. b) Entropy rate $h(\tau)$ of the chaotic laser system for a fixed $N_{\rm\epsilon}=3$ bits. The $h_0$ limit (Eq. \ref{eq:h0}) is the information theoretical limit for the entropy rate given the PDF and bandwidth of the signal. Since we do not have a theoretical prediction for the PDF of the laser chaos, we estimate the PDF using the experimental histogram shown in Fig. \ref{fig:laserapparatus} and determine $D_{KL}=-0.95$ bits. $h_{CP}$ is the Cohen-Procaccia entropy rate estimate performed on the data, as described in the Appendix. Here we show $h_{CP}$ using embedding dimensions (pattern lengths) of $d=1$ and $d=6$. MCV=$-\tau^{-1}\log_2(p_{max})$ is the Most Common Value entropy rate estimate from the NIST draft recommendations \cite{90B}. The Markov estimate, also from the NIST draft recommendations, takes into account first-order correlations in the data. $h_{CP}$ ($d=3$) for the electronic noise in the detectors is also shown.}
\label{fig:laserentropy}
\end{figure}

Figures \ref{fig:laserentropy}a and \ref{fig:laserentropy}b compare the results of different entropy rate estimates on the chaotic laser signal. For the Cohen-Procaccia estimate, we use $d=6$ embedding dimensions. While it is unlikely that the attractor can be fully embedded in 6 dimensions, we choose $d=6$ because the entropy rate estimate did not change much for $d>6$. The bandwidth used in determining $h_0$ is the detector bandwidth $BW=33$ GHz because the bandwidth of the chaos (34 GHz) is greater than the detector bandwidth. Because we do not have a theoretical prediction for a PDF for the chaotic laser system, we use the experimentally measured PDF for computing $D_{KL}$ in Eq. \ref{eq:h0}. 

In Fig. \ref{fig:laserentropy}a, we consider $h(\epsilon)$ for a fixed $\tau$. $N_{\rm \epsilon}$ is the number of bits per sample measured by the detector. For the NIST tests $N_{\rm \epsilon}$ is obtained by taking the appropriate number of most significant bits from an 8-bit oscilloscope measurement; for the Cohen-Procaccia estimate $N_{\rm \epsilon}$ is obtained by referencing the bin width $\epsilon$ (described in section \ref{sec:cp}) to the full 8-bit resolution of the oscilloscope (1.6mV). We also show the thresholding ($N_{\rm\epsilon}=1$ bit) limit $h = 2BW$ bits/s as a blue dotted line. The shading above these limits denotes a region of unobtainable entropy rates for a system with the experimental PDF shown in Fig. \ref{fig:laserapparatus}c. For comparison, we show the $d=1$ Cohen-Procaccia entropy rate estimate, which estimates the entropy of the experimentally measured histogram and considers no temporal correlations. This shows what the entropy rate would be if the system were actually IID when sampled at $\tau^{-1}=$100 GSamples/s. Of course, the system \textit{cannot} be IID when sampled at faster than 66 GSamples per second, since the detector bandwidth is 33 GHz. Unsurprisingly, both the MCV and the $d=1$ estimates significantly overestimate the entropy rate. The Markov estimate does little better, still providing an entropy rate estimate that is significantly higher than the $h_0$ limit. The $d=6$ Cohen-Procaccia estimate agrees well with the Eq. (2) limit for $N_{\rm\epsilon}\geq2$. However, the Cohen-Procaccia estimate is unable to provide a $d=6$ estimate for $N_{\rm\epsilon}>4$ due to the data requirements (we used 1GB of data). 

Fig. \ref{fig:laserentropy}a also shows the entropy rate obtained by simply turning off all optics and measuring the electronic noise at the detector. It is clear that the electronic noise in the detector contributes a significant fraction (about 20\%) of the entropy at full resolution.

Fig. \ref{fig:laserentropy}b shows the entropy rate as a function of the sampling rate $\tau^{-1}$ for a fixed $N_{\rm\epsilon}=3$ bits. As the sampling rate is increased, the maximum entropy rate increases, then starts to plateau at a sampling rate of about 50 GSamples per second. As expected, the MCV estimate detects no correlations and continues to increase for $\tau^{-1} > 2$BW; the Markov estimate does only a little better, showing a slight roll off. The $d=6$ Cohen-Procaccia estimate has the most noticeable roll off, indicating that it detects temporal correlations better than the other two methods. By detecting these correlations, the Cohen-Procaccia algorithm informs us that the experiment is not purely random at high sampling rates; this can inform the choice of $\tau^{-1}$ in RNG design.  For example, it suggests that this system should be sampled at $\tau^{-1}<50$ GSamples/s if the designer wants there to be no temporal correlations.

As noted above, there are two important limits to consider in the design of a RNG based on chaotic lasers: $h_{KS}$ and $h_0$. One might wonder about the interplay between $h_{KS}$, which describes the dynamics of the chaotic system and $h_{SH}$, which limits the amount of information that can be transmitted through a finite bandwidth channel. It has long been known that filtering a chaotic system does not change the $h_{KS}$ \cite{badii1988dimension,mitschke1988measuring}. Passing a chaotic signal through a linear filter simply makes the current output of the filter some linear combination of all the previous inputs to the filter. In principle, if one knows the linear combination that describes the filter, one can determine the unfiltered output of the chaotic system from the filtered output and thus obtain an entropy rate of $h_{KS}$. 

It might seem, then, that a RNG based on a chaotic system can violate the Shannon-Hartley limit described above by low-pass filtering a chaotic signal so that $h_{SH}<h_{KS}$. This is not the case. There is a minimum resolution necessary to obtain $h_{KS}$, as described in the Appendix and ref. \cite{gaspard1993noise}. One must use a higher resolution to obtain $h_{KS}$ from the filtered chaotic system than is necessary to obtain $h_{KS}$ from the unfiltered system \cite{kantz2004fast}. This increase in resolution increases $N_{\rm\epsilon}$ in such a way as to cancel the decrease in bandwidth $BW$ and ensure that the Shannon-Hartley limit (Eq. \ref{eq:shannonhartley}) is not violated.

Essentially, $h_{KS}$ describes the rate at which the chaotic system generates entropy. This is a property of the physical entropy source and is independent of filtering or any other part of the digitization process. $h_{SH}$ and $h_0$ describe the rate at which entropy can be harvested by the measurement apparatus given the resolution and bandwidth limitations of the measurement apparatus. In short, just because a physical system is generating entropy at a rate $h_{KS}$ does not mean that a given measurement apparatus is able to harvest that much entropy from the system.

$h_{KS}$ can in principle be determined from a deterministic model of the chaotic system by calculating the Lyapunov spectrum; however, a reliable estimate of $h_{KS}$ from experimental data is problematic due to the high dimensionality of three time-delayed chaotic lasers. We note that the largest Lyapunov exponent has been calculated numerically to be on the order of several ns$^{-1}$ for two cascaded chaotic lasers \cite{kanno2016complexity}. We expect that $h_{KS}$ should be several times greater than this for our three cascaded laser system, since the additional chaotic laser should increase the complexity and $h_{KS}$ depends on all the Lyapunov exponents, not just the largest one. Our entropy estimates shown in Fig. 6 are consistent with this expectation.

\subsection{RNG from Amplified Spontaneous Emission}
 The final optical RNG technique we analyze is the detection and digitization of optically filtered amplified spontaneous emission (ASE) noise from a light source such as superluminescent diode (SLD). ASE sources provide an easily measurable, high bandwidth noise signal and have been used for RNG since 2010 \cite{williams2010fast}. Because it is inherently quantum mechanical in origin, ASE cannot be described deterministically; thus, entropy can be harvested by detecting and sampling the ASE signal.

\begin{figure}
\centering
\includegraphics[width=\textwidth]{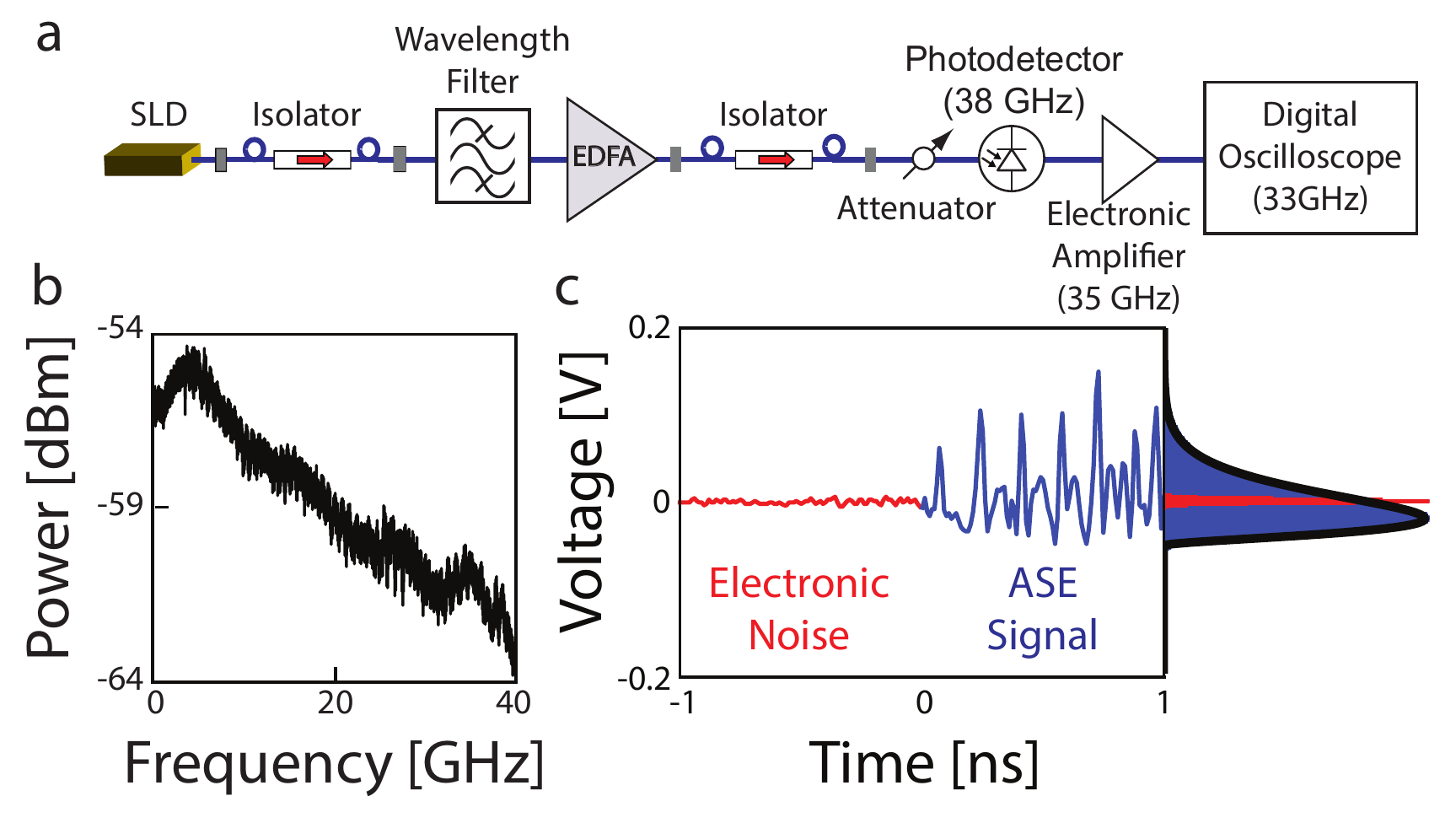}

\caption{a) Experimental setup for digitizing filtered ASE signal from a SLD. The injection current of the SLD is 300.0 mA. b) RF spectrum for detected ASE signal when the optical filter has a width of 0.6nm and center frequency of 1550nm. The 90\% bandwidth of the resulting signal is 31 GHz. c) Time series and PDFs of electronic noise only (red) and detected ASE signal (blue) when optical filter has a width of 0.6nm. The solid black line is the best-fit of Eq. \ref{eq:gamma} to the experimental data.}
\label{fig:SLDapparatus}
\end{figure}

There have been several different but closely related schemes to generate random numbers from ASE sources \cite{williams2010fast,li2011scalable,argyris2012sub,wei2012high,yamazaki2013performance,li2014random,hughes2016strengthening,zhang2016fast}. Here, we discuss the system depicted in Fig. \ref{fig:SLDapparatus}a. The ASE output of a SLD (DenseLight Semiconductors DL-CS5254A-FP) passes through an optical isolator, a tunable optical filter (Santec OTF-970), and an erbium-doped fiber amplifier (EDFA, PriTel FA-18-IO). This optical intensity is then attenuated before being detected by a 38 GHz photodetector (New Focus 1474-A). The electrical signal from the photodetector is amplified by a 35 GHz electronic amplifier (Picosecond pulse labs, 5882-219) then sampled by a high speed 8-bit oscilloscope (Tektronix DPO73304D, 33 GHz bandwidth, 100 GigaSamples/s). 

\begin{figure}
\centering
\includegraphics[width=.7\textwidth]{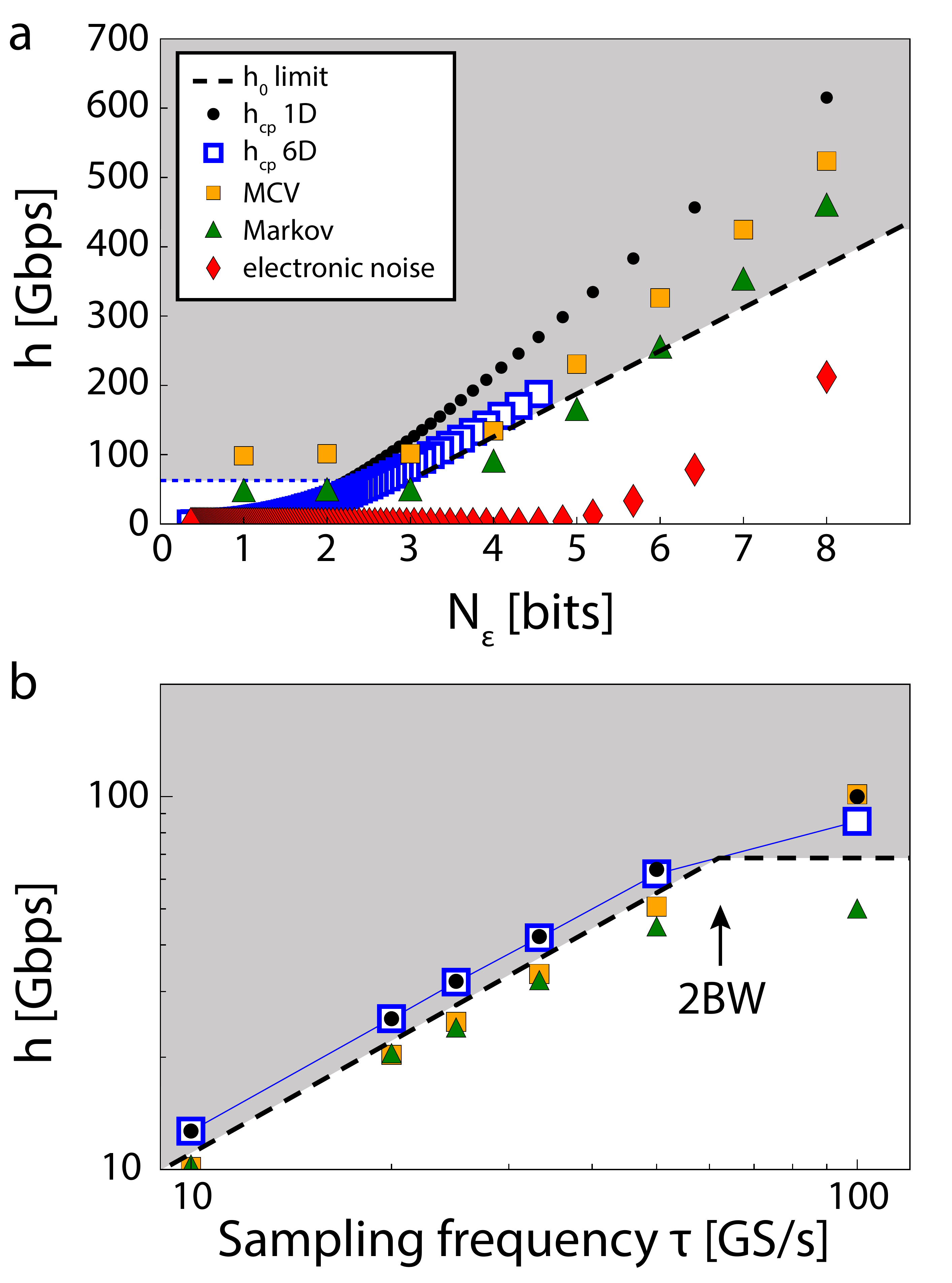}
\caption{a) Entropy rate $h(\epsilon)$ of the ASE signal from an SLD and a 0.6nm filter for a fixed $\tau^{-1}=100$ GSamples/s. The blue dotted line is the entropy rate obtained by thresholding at the median, $\tau^{-1}$ b) Entropy rate $h(\tau)$ from an SLD and a 0.6nm filter for a fixed $N_{\rm\epsilon}=3$ bits. The $h_0$ limit (Eq. \ref{eq:h0}) is the information theoretical limit for the entropy rate given the PDF and bandwidth of the signal ($D_{KL}=-1.94$ bits for the best-fit PDF from Eq. \ref{eq:gamma} and shown in Fig. \ref{fig:SLDapparatus}). $h_{CP}$ is the Cohen-Procaccia entropy rate estimate performed on the data, as described in the Appendix. Here we show $h_{CP}$ using embedding dimensions (pattern lengths) of $d=1$ and $d=6$. MCV=$-\tau^{-1}\log_2(p_{max})$ is the Most Common Value entropy rate estimate from the NIST draft recommendations \cite{90B}. The Markov estimate, also from the NIST draft recommendations, takes into account first-order correlations in the data. $h_{CP}$ ($d=6$) for the electronic noise in the detectors is also shown. }
\label{fig:SLDentropy}
\end{figure}

The optical filter is used to control the bandwidth of the ASE signal. It has been shown that the optimal optical filter bandwidth for RNG is approximately equal to the photodetector bandwidth \cite{williams2010fast}. In this case, we used an nominal optical bandwidth of 0.6nm at 1550 nm center wavelength. The experimentally measured power spectrum of the ASE signal with this filter is given in Fig. \ref{fig:SLDapparatus}b; the 90\% signal bandwidth is 31 GHz. The PDFs and representative time series of both the full ASE signal and the electronic noise (with all optics turned off) are given in Fig. \ref{fig:SLDapparatus}c. As in the case of the chaotic laser, one can tell from the time series in Fig. \ref{fig:SLDapparatus}c that a few of the least significant bits of the signal are due to electronic noise rather than the optical signal. This is confirmed by the entropy analysis in Fig. \ref{fig:SLDentropy}a.

While the probability distribution of the photocurrent output by the photodetector depends on the properties of the optical filter and photodetector used, it is known that the photocurrent distribution can be reasonably approximated by the gamma distribution

\begin{equation}
\label{eq:gamma}
p_i(x)=\frac{x^{a-1}\exp{[-x/b]}}{b^a\Gamma(a)},\qquad x>0
\end{equation}
where $a$ is the shape parameter (signal-to-noise ratio) and $b$ is the scale parameter \cite{williams2010fast,goodman1985statistical}. Fig. \ref{fig:SLDapparatus}c shows that this is a reasonablly good approximation in this case, with $a$=2.77 and $b$=17.7 mV. 

Figures \ref{fig:SLDentropy}a and \ref{fig:SLDentropy}b compare the results of different entropy rate estimates on the ASE signal. As with the chaotic laser, $N_{\rm \epsilon}\leq8$ is obtained by taking the appropriate number of most significant bits from an 8-bit oscilloscope measurement for the NIST tests; for the Cohen-Procaccia estimate $N_{\rm \epsilon}$ is obtained by referencing the bin width $\epsilon$ to the full 8-bit resolution of the oscilloscope (1.6mV). The upper limit $h_0$ is given by Eq. \ref{eq:h0} as a black dashed line and the thresholding ($N_{\rm\epsilon}=1$ bit) limit $h = 2BW$ bits/s as a blue dotted line. Here, $BW=31$ GHz since the 90\% signal bandwidth is the smallest relevant bandwidth. We use Eq. \ref{eq:gamma} with best-fit parameters $a$=2.77 and $b$=17.7 mV to determine $D_{KL}$ and $h_0$. The shading above these limits denotes a region of unobtainable entropy rates for a system with the PDF given by Eq. \ref{eq:gamma}. In Fig. \ref{fig:SLDentropy}a, the $d=1$ Cohen-Procaccia estimate shows what the entropy rate would be if the system were IID when sampled at $\tau^{-1}=100$ GSamples/s. Of course, the system \textit{cannot} be IID when sampled at faster than 62 GSamples per second, since the signal bandwidth is 31 GHz. Thus, it is expected that the $d=1$ Cohen-Procaccia estimate would lie in the shaded region of unobtainable entropy rates.

For this data, we use the Cohen-Procaccia entropy rate estimate with $d=6$ dimensions because the entropy rate estimate did not change much for $d>6$. The $d=6$ Cohen-Procaccia estimate closely follows the $h_0$ limit, while the other entropy estimates show a significantly larger slope, resulting in large overestimates of the entropy rate. The $d=6$ Cohen-Procaccia lies slightly above the $h_0$ limit; this is likely due to the mismatch between the theoretical PDF in Eq. \ref{eq:gamma} and the actual experimental PDF. The $d=6$ Cohen-Procaccia estimate is the only one of the 4 entropy estimates shown that seems to follow the $h_0$ limit; however, the Cohen-Procaccia estimate is unable to provide a $d=6$ estimate for $N_{\rm\epsilon}>4$ due to the data requirements. 

Fig. \ref{fig:SLDentropy}a also shows the entropy rate obtained by simply turning off all optics and measuring the electronic noise at the detector. As for the case of the chaotic laser, it is clear that the electronic measurement noise contributes a significant fraction (about 20\%) of the entropy at full resolution.

Fig. \ref{fig:SLDentropy}b shows the entropy rate as a function of the sampling rate $\tau^{-1}$ for a fixed $N_{\rm\epsilon}=3$ bits. Again, the dashed black line denotes the upper limit provided by Eq. \ref{eq:h0}. For lower sampling rates, the Cohen-Procaccia estimate is slightly above the Eq. \ref{eq:h0} limit; this is due to the mismatch between the theoretical and experimental PDFs shown in Fig. \ref{fig:SLDapparatus}c. As the sampling rate is increased, the maximum entropy rate increases, then plateaus as the sampling rate approaches twice the signal bandwidth (62 GSamples/s). As expected, the MCV estimate detects no correlations and continues to increase for $\tau^{-1} > 2$BW; The Markov and Cohen-Procaccia estimates do quite a bit better, showing a roll off with increasing sampling rate. This indicates that there are temporal correlations in the data. By detecting these correlations, the entropy estimates inform us that the experiment is not behaving purely stochastically for sampling rates that are too high. While the Markov and Cohen-Procaccia estimates perform similarly for $N_{\rm\epsilon}=3$, it is clear from Fig. \ref{fig:SLDentropy}a that the Markov estimate does not give valid results for $N_{\rm\epsilon}\geq7$.

Since ASE sources can have large bandwidth, the main factor that limits the entropy rate for ASE sources is the bandwidth of the measurement apparatus. Thus, as the bandwidths of photodetectors and digitizers improve, we expect the entropy rate of ASE sources to similarly increase. Additionally, a single ASE source can be used to generate multiple independent bitstreams by taking different slices of the optical spectrum, as done in ref. \cite{li2011scalable}. 

\section{\label{sec:conclusion}Conclusions}
Physical RNGs are becoming increasingly important in digital communications and security, as evidenced by their widespread commercial availability, both embedded in CPUs \cite{hamburg2012analysis} and as external devices \cite{hughes2016strengthening,picoquant,IDQuantique}. Optical and photonic systems are leading the way as physical sources of randomness due to their high speed, access to the inherent randomness in quantum mechanical phenomena, and resistance to external interference from electric and magnetic fields. In the last decade, optical RNGs have elevated the state-of-the-art from a few hundred Mbits/s to one Tbit/s.

Motivated by this race for the highest random bit rates, researchers have often been insufficiently concerned about where the entropy is coming from. Instead, the standard practice is to sufficiently post-process some unpredictable signal so that the final output bit sequence can pass statistical tests designed for PRNGs. As we discussed above, the fastest physical entropy sources involve the digitization of high-bandwidth, unpredictable analog waveforms. The digitization process naturally forces one to think about what the measurement resolution $\epsilon$ and sampling rate $\tau^{-1}$ should be. As we have shown, the choices of $\epsilon$, $\tau$, and post-processing technique can determine which physical process or processes contribute to the extracted entropy rate.

The new NIST draft recommendations for the evaluation of physical RNGs \cite{90B} come a long way; they suggest that one estimate the entropy using minimally post-processed data and require some physical justification of where the entropy is coming from. However, the new standards do not recognize dynamical entropy sources or the importance of the digitization process ($\epsilon$ and $\tau$).

We recommend that physical RNG evaluation techniques evolve away from statistical tests designed for PRNGs toward entropy estimates that provide insight into the physical origins and limitations of the optical entropy source. In order to acheive this, we recommend that RNG designers perform an $(\epsilon,\tau)$ entropy analysis on the raw digitized data (as in Fig. 1b) using a variety of entropy estimates, including the the Cohen-Procaccia estimate and tests from the NIST entropy estimation suite. The $h(\epsilon,\tau)$ analysis, in conjunction with considering simple physical and information theoretical limits of entropy generation, provides more than a simple pass/fail validation of a RNG; it provides relevant information about an entropy source such as how finely and frequently to sample the source and what types of post-processing and conditioning should be used to extract entropy from the desired source. 

As an example, we have performed this analysis for three state-of-the-art optical entropy sources. We found that the digitization of unpredictable, high-bandwidth analog signals generates significantly higher entropy rates than does single photon detection. Chaotic lasers and ASE signals can produce similar entropy rates (on the order of hundreds of Gbits per second); however, the simplicity of the ASE setup is attractive as is the ability for one SLD to generate multiple independent bitstreams, as done in ref. \cite{li2011scalable}. We also found that post-processing methods that use least-significant bit extraction might be taking their entropy from electronic noise in the detector or digitizer rather than from the desired optical entropy source.

\section{A look to the future}
One of the most important areas for future research in random number generation is the one with which this Perspective is most concerned: evaluation standards and practices. It is the opinion of the authors that the standards for physical RNGs should involve entropy estimates of the raw physical data and that the Cohen-Procaccia entropy rate estimate is one useful entropy measure; however, there is no single perfect entropy rate estimate. The upcoming release of the official NIST recommendations for physical entropy sources will provide some new entropy estimation techniques. These will need to be tested on all types of optical entropy sources. The statistical analysis of data from physical entropy sources is a highly challenging but important problem in the field of RNG. We hope that this Perspective as well as the upcoming release of the NIST recommendations will lead the optical RNG community to embrace entropy estimation from the raw physical data, and to continue to develop new and better entropy estimates.

Another crucial issue facing the optical RNG community is to bridge the gap between the ultrafast RNG rates possible in the lab ($\sim$1 Tb/s) and the significantly slower speed of commercially available optical RNGs ($\sim$1 Gb/s). The ultrafast rates in laboratory experiments have been obtained by taking one-time measurements with an oscilloscope; they cannot be sustained for more than tiny fractions of a second. Further, the postprocessing necessary to extract the entropy is often done offline. One critical path of future research is developing real-time implementations of the post-processing necessary for optical RNGs. There has been some work in this area \cite{ugajin2017real,wang20134, abellan2015generation, shinohara2017chaotic}; however, as of now real-time implementations of optical RNGs have a long way to go to read the Tb/s rates of the one-time oscilloscope measurements. Once post-processed, there is the problem of transferring the data to the memory of the user in real time. High-speed entropy extraction and data transfer are two major practical problems that the field of optical random number generation will have to address in the coming years.

Additionally, in order to be practically useful, the size and cost of the laboratory RNGs must be reduced. One promising way to do this is by implementing the optical RNG on photonic integrated circuits \cite{ugajin2017real}. Low-cost photonic integrated circuits are currently being developed for optical computing and information transfer within traditional computing; we anticipate that optical RNGs will begin to be developed on chip as well. These circuits provide the additional benefit that they are robust against temperature fluctuations and air turbulence.

We also anticipate the development of optical RNGs with special properties suited for application-specific purposes. One recent example of this is the laser phase noise-based RNG used in the recent loophole-free Bell tests \cite{shalm2015strong, hensen2015loophole, giustina2015significant}. Each of these tests relied on the real-time optical RNG described in ref. \cite{abellan2015generation} to randomly and independently choose the measurement bases such that the choice of basis is space-like separated. This allowed the researchers to close the locality loophole.

Over the last decade, much of the optical RNG research has focused on breaking bit rate records that pass the statistical test suites. Once the optical RNG community shifts focus from record breaking to entropy analysis and physical origins of randomness, we will also see an increasing focus on decreasing the size and power constraints and increasing the robustness to external (potentially unsafe) noise sources. When all of these considerations are taken into account, optics and photonics will emerge as the most promising technology for physical random number generation.

\section*{\label{sec:cp}Appendix}
\subsection*{A comment about the relationship between PRNGs and deterministic chaos}

We state above that PRNGs cannot increase the entropy rate of their input because they are deterministic algorithms. Given the set of equations and the initial conditions, one can calculate the full future output of the system. However, it is well-established that deterministic chaotic systems do have an associated entropy rate, $h_{KS}$ \cite{ott2002chaos,boffetta2002predictability,gaspard1993noise}. One might might wonder why a chaotic system can generate entropy while a PRNG cannot. 

PRNGs require a finite-length seed (initial condition) as input, which contains all the entropy \cite{90A}. The PRNG then performs deterministic computations on the seed in order to generate pseudo-random numbers as output. These numbers appear random: they are uniformly distributed and pass all statistical tests of randomness. The numbers are called pseudo-random because for a given seed, the PRNG will always produce the exact same output sequence. Therefore, if an attacker obtains the seed, the future output of the PRNG is completely predictable. 

Due to the finite precision of computers, a PRNG with a given seed will repeat after a finite (often very large) number of iterations \cite{falcioni2005properties}; that is, once the entropy from the seed is used up, no new entropy can be obtained from the PRNG. To resolve this, PRNGs are often re-seeded somewhat frequently with additional entropy. Essentially, then, the PRNG serves to reveal the entropy from the seed at a given rate and with some desired properties (e.g. uniformly distributed output); however, the entropy itself must be provided from some other source (often physical entropy sources). Due to finite precision constraints, computer simulation of a deterministic chaotic system is essentially a PRNG and will encounter these same restrictions \cite{falcioni2005properties}.

We now consider mathematical chaotic systems that are described by deterministic equations with infinite precision. As in the case of PRNGs on a classical computer with finite precision, the entropy in an infinite-precision chaotic system is stored in the initial condition. Unlike in a PRNG, in an infinite-precision chaotic system, an infinite amount of information (or entropy) is stored in the infinitely precise initial condition. Imagine that an observer of this system cannot measure the initial condition to more than a handful of most significant bits. The chaotic system, due to its inherent sensitivity to initial conditions, amplifies the bits of lower significance so that they become measurable \cite{grassberger1985information, shaw1981strange}. The average rate at which this information about the precise value of the initial condition is revealed by the chaotic system can be quantified by the Kolmogorov-Sinai entropy rate, which is equal to the sum of the positive Lyapunov exponents \cite{gaspard1993noise,boffetta2002predictability}.

Of course, infinite-precision chaotic systems do not exist in the real world; all physical systems are at some level granular, quantized, and susceptible to sources of noise and uncertainty, which together prevent the physical chaotic system from having infinitely precise initial conditions. This uncertainty continuously scrambles the least significant bits of the state of the system, continuously re-seeding the chaotic system. The noise is amplified by the chaos \cite{fox1991amplification, bracikowski1992amplification} and contributes to the entropy production. In the case of laser chaos, the intrinsic noise has been considered to be quantum mechanical in origin and due to spontaneous emission in the laser \cite{sunada2012noise}.”

There may also be classical sources of noise. These may be intrinsic, in which case they will be amplified by the chaos, or external (e.g. measurement noise), in which case they may or may not contribute to the measured entropy rate, depending on the $\epsilon$ and $\tau$ and post-processing used in the physical RNG, as discussed above.

In summary, for both PRNGs and ideal chaotic systems, the entropy comes from the initial conditions; PRNGs and chaotic systems are similar in that they amplify the bits of low significance of their initial conditions. The fundamental difference is that the amount of entropy in the initial conditions of a PRNG is limited by the finite precision of a computer, while the initial conditions of an ideal chaotic system has infinite precision. Because of the finite precision, PRNGs must eventually repeat and therefore are periodic systems with long periods; however, chaotic systems never repeat. Once a PRNG repeats, it has used up all the entropy in its seed and the output entropy rate is 0; a chaotic system reveals the infinitely many low-significance digits of the initial conditions at an average rate of $h_{KS}$ forever. Further, because of finite measurement precision and intrinsic noise, chaotic systems are truly random; Gaspard and Wang show that a chaotic system and a stochastic Markov chain with the same $h_{KS}$ have the same degree of dynamical randomness \cite{gaspard1993noise}.

\subsection*{The $h_0$ limit}
In Eq. \ref{eq:h0} we provide the information theoretical limit for the maximum entropy that can be harvested with a given probability distribution $p(x)$ with a signal and measurement bandwidth of $BW$. Here we derive this from a previous result and explain how to calculate it in practice.

Gaspard and Wang \cite{gaspard1993noise} give the upper limit of $\epsilon$-entropy per sample as 
\begin{equation}
H_0(\epsilon)=-\log_2(\epsilon)-\int \diff x\, p(x)\log_2p(x)+O(\epsilon),
\end{equation}
where $p(x)$ is the PDF of the signal from which entropy is being harvested. One can either use a theoretical PDF or estimate the PDF from an experimentally measured histogram for $p(x)$. A theoretical PDF is preferable, since one can calculate the integral exactly.
We also know that the sampling rate is limited by information theory to the Nyquist rate $f_{max}=2BW$. Here, $BW$ is the limiting bandwidth, which is the minimum of all relevant bandwidths (signal bandwidth, detector bandwidth, digitizer bandwidth, etc.). Thus, as $\epsilon\to 0$
\begin{equation}
\label{eq:h0calc}
h_0(\epsilon)=\min(\tau^{-1},2BW)\big(-\log_2(\epsilon)-\int \diff x\, p(x)\log_2p(x)\big).
\end{equation}
In practice, it is often easiest to use Eq. \ref{eq:h0calc} to determine the $h_0$ limit. However, the limit can be understood intuitively by writing it in terms of $N_{\rm\epsilon}$ and the Kullback-Leibler divergence, as we now show.

The number of bits per sample $N_{\rm\epsilon}=\log_2(\frac{b-a}{\epsilon})$, where $a$ and $b$ are the end points of the measurement range of the digitizer. If we define $U=\frac{1}{b-a}$, we can write $N_{\rm\epsilon}=-\log_2(U\epsilon)$ and
\begin{equation}
\label{eq:h0derive}
h_0(\epsilon)=\min(\tau^{-1},2BW)\big(N_{\rm\epsilon}+\log_2U-\int \diff x\, p(x)\log_2p(x)\big).
\end{equation}
Now define $u(x)$=U for $a\leq x\leq b$ and $u(x)=0$ for all other $x$. $p(x)$ is also only non-zero for $a\leq x\leq b$. Since $\int p(x)dx$=1, we can write $\log_2U=\int p(x)\log_2U\diff x$. We can then combine the integrals:

\begin{equation}
h_0(\epsilon)=\min(\tau^{-1},2BW)\big(N_{\rm\epsilon}-\int \diff x\, p(x)\log_2(\frac{p(x)}{u(x)})\big).
\end{equation}

The Kullback-Leibler divergence is defined as $D_{KL}(p(x)||u(x))\equiv-\int \diff x\, p(x)\log_2(\frac{p(x)}{u(x)})$, giving us Eq. \ref{eq:h0}.

\subsection*{Cohen-Procaccia entropy}
In this section, we present the Cohen-Procaccia algorithm as a method to estimate $h(\epsilon,\tau)$ \cite{cohen1985computing}. The Cohen-Procaccia estimate is especially useful because it can be used to directly compare stochastic and chaotic sources \cite{gaspard1993noise}. Further, it does not unnecessarily penalize entropy sources with PDFs that are approximately normal like some of the NIST draft tests do \cite{kelsey2015predictive}.

\begin{figure}
\centering
\includegraphics[width=0.95\textwidth]{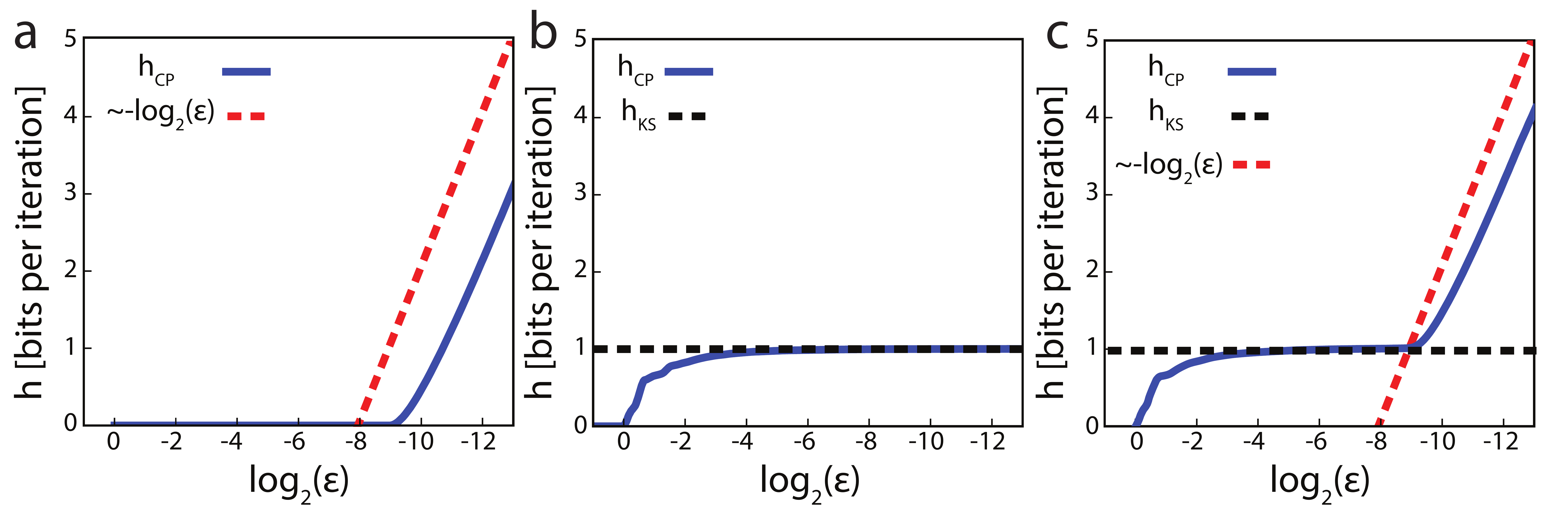}
\caption{\label{fig:hCP}Demonstration of the Cohen-Procaccia entropy rate 
estimate (blue line) on a simple time series. \textit{a} Gaussian noise with strength $a$=0.001. \textit{b} Logistic map with parameter $r$=4. \textit{c} Logistic map with additive Gaussian noise with standard deviation 0.001. The red dashed line shows a line with slope $-\log_2(\epsilon)$. The black dashed line gives the Kolmogorov-Sinai entropy rate of the logistic map with $r$=4 ($h_{KS}$=1 bit per sample).}
\end{figure}

The Shannon entropy, or average amount of information contained per sample, of a random variable $X$ is given by
\begin{equation}
\label{eq:shannon}
H(X)=-\sum p(x)\log_2(p(x)),
\end{equation}
where the summation is taken over all possible values of $x$ \cite{coverthomas}.
For a joint probability distribution, this definition of entropy extends naturally to
\begin{equation}
H(X_1,X_2,...,X_d)=-\sum p(x_1,x_2,...,x_d)\log_2(p(x_1,x_2,...,x_d)).
\end{equation}

One definition of entropy rate is the average amount of new information obtained by measuring the current sample given the history of previous samples:

\begin{equation}
\label{eq:entropyrate}
h=\lim_{d\to\infty}\frac{1}{d\tau}H(X_d|X_{d-1},X_{d-2},...,X_{1})=\lim_{d\to\infty}\frac{1}{\tau}\big(H(X_d,X_{d-1},...,X_1)-H(X_{d-1},X_{d-1},...,X_1)\big),
\end{equation}
where $\tau^{-1}$ is the rate at which the distribution is sampled. In Eq. \ref{eq:entropyrate}, one considers the rate at which the entropy of the set of patterns of length $d$ symbols changes with $d$. 

To calculate the entropy of a dynamical system, the patterns of length $d$ that are used are obtained by a $d$-dimensional time-delay embedding \cite{ott2002chaos} of the data with delay $\tau$. The time-delay vectors can be considered samples from a $d$-dimensional probability distribution in phase space. $h(\epsilon,\tau)$ can then be calculated according to Eq. \ref{eq:entropyrate}. 

For an IID random process, each sample will be completely independent of all previous samples, so Eq. \ref{eq:entropyrate} becomes $h=\tau^{-1}H(X_d)$. However, when there are temporal correlations of length $d\tau$ or less, the $d$-dimensional pattern entropy rate will be reduced. 

In principle, one can estimate the Shannon entropy directly. First estimate the $d$-dimensional joint probability distribution by making a histogram with $d$-dimensional boxes of width $\epsilon$ and use this in Eq. \ref{eq:entropyrate} to estimate the entropy rate. This approach requires a large amount of data and computing resources for systems with large embedding dimension.

Cohen and Procaccia \cite{cohen1985computing} developed a more efficient way to estimate the entropy rate in order to estimate the Kolmogorov-Sinai entropy from experimental data of chaotic systems. For additional information about the close relationship between the Kolmogorov-Sinai entropy and the Shannon entropy, see ref. \cite{boffetta2002predictability}. Gaspard and Wang \cite{gaspard1993noise} later showed that the Cohen-Procaccia algorithm can also be accurately estimate the entropy rate of stochastic systems. We now briefly review the Cohen-Procaccia algorithm.

First, one makes the previously described $d$-dimensional time-delay embedding. Then one randomly selects $M$ of these points as reference points. Typically $M$ is much smaller than the length of the time series. For each reference point $n$, one computes $f_n(\epsilon)$, the fraction of other points within a $d$-dimensional box of width $\epsilon$ (that is, within a distance $\epsilon/2$ of the reference point). Here distance is given by the square metric dist[$\mathbf{x},\mathbf{y}$]$=\max\{|x_1-y_1|,|x_2-y_2|,...,|x_d-y_d|\}$, where $\mathbf{x}$ and $\mathbf{y}$ are two $d$-dimensional vectors. The $d$-dimensional pattern entropy estimate is then given by

\begin{equation}
\label{eq:cppatternentropy}
H_d=-\frac{1}{M}\sum_{n=1}^M\log_2(f_n(\epsilon)).
\end{equation}

The Cohen-Procaccia entropy rate estimate is then obtained by using Eq. \ref{eq:cppatternentropy} in Eq. \ref{eq:entropyrate}

\begin{equation}
h_{CP}(\epsilon,\tau,d)=\tau^{-1}(H_d(\epsilon,\tau)-H_{d-1}(\epsilon,\tau)),
\end{equation}
where we have explicitly added in the dependence of $h$ and $H$ on the box width $\epsilon$ and the embedding time-delay $\tau$. The only differences between the Cohen-Procaccia estimate and a direct calculation of the Shannon entropy are that the Cohen-Procaccia estimate uses $M$ reference points, and that the histogram bins are centered on the reference points instead of being a fixed rectangular array. The Cohen-Procaccia calculation still requires a large amount of data, but is much more computationally efficient than a direct calculation of the Shannon entropy. We note that placing the centers of the bins on the reference points results in poor entropy rate estimates for large $\epsilon$ (small $N_{\rm\epsilon}$), but accurate estimates for small $\epsilon$.

Gaspard and Wang \cite{gaspard1993noise} used the Cohen-Procaccia estimate to compare the entropy generation rates of Gaussian noise and the logistic map $x_{n+1}=rx_{n}(1-x_n)$ with $r$=4. We have replicated these results and present them in Fig. \ref{fig:hCP}a and b. The Cohen-Procaccia estimate can distinguish a stochastic process from a chaotic one by revealing the dependence of the entropy rate on $\epsilon$. For a stochastic process, $h(\epsilon)$ should scale with $\log(\frac{1}{\epsilon})$, while $h(\epsilon)$ should converge to $h_{KS}$ as $\epsilon\to0$ for a chaotic process. This behavior is captured by the Cohen-Procaccia algorithm for the logistic map in Figs. \ref{fig:hCP}a and \ref{fig:hCP}b. In Fig. \ref{fig:hCP}b, for low resolution (large $\epsilon$), no entropy can be harvested from the system because the resolution is too coarse. As the resolution increases, the amount of entropy that can be extracted from the system also increases, until the full attractor is resolved. At this point, $h_{CP}(\epsilon)$ plateaus at $h_{KS}$, even as $\epsilon\to0$.

Gaspard and Wang also considered a noisy logistic map, in which the time series was obtained by iterating the logistic map with $r$=4 then adding Gaussian noise of standard deviation 0.001 to the output. This result is shown in Fig. \ref{fig:hCP}c. For low resolution (large $\epsilon$), the entropy increases as the resolution increases. At intermediate resolution, the chaotic attractor can be fully resolved, but the noise cannot, so $h(\epsilon)$ plateaus at $h_{KS}$. At high resolution (small $\epsilon$), the noise is resolved and $h(\epsilon)$ scales as $\log(\frac{1}{\epsilon})$. It is clear from the figure not only that the Cohen-Procaccia estimate can accurately predict the rate of entropy production of both stochastic and chaotic systems, but also that the scaling of the entropy rate with $\epsilon$ can provide some information about where the entropy is coming from at a given measurement resolution.

As we mentioned previously, an ($\epsilon,\tau$) entropy rate analysis can be performed with any entropy estimation method. We use the Cohen-Procaccia estimate because it is known to treat both chaotic and stochastic sources fairly \cite{gaspard1993noise,hagerstrom2015harvesting} and it can recognize higher order correlations than metrics such as the MCV and Markov estimates.
While in principle, the Cohen-Procaccia algorithm can identify correlations of any time scale, this requires an impractically large amount of data if the time scale of the slowest correlations is much slower than the fastest time scale. We do not consider this a significant problem, however, because good physical RNG design involves stabilization techniques to remove slow fluctuations due to external factors, such as power supply or temperature fluctuations.

However, there are some disadvantages to the Cohen-Procaccia entropy rate estimate. In general, the Cohen-Procaccia algorithm does require significantly more data than the NIST entropy estimation suite does, and the amount of data needed increases with the dimension of the entropy estimate. It does, however, pick up higher order correlations than do the NIST entropy estimation techniques. As a result, for entropy analyses of systems that require a high dimension for the Cohen-Procaccia algorithm to converge, it may be difficult to obtain an accurate estimate of the entropy rate. Still, the Cohen-Procaccia algorithm has value: for example, if the entropy rate estimate varies with dimension, then it is clear that the system is not behaving completely randomly and that the entropy source should probably be sampled less frequently. Once the Cohen-Procaccia entropy rate estimate does converge, then one has both an entropy rate estimate and some assurance that they are sampling at an appropriate rate.

\subsection*{Acknowledgments} JDH, TEM, and RR thank the U.S. Department of Defense for support. JDH and RR acknowledge support from ONR grant N000141612481 AU and YT acknowledge support from Grants-in-Aid for Scientific Research from the Japan Society for the Promotion of Science (JSPS KAKENHI Grant Number JP16H03878) and from Management Expenses Grants from the Ministry of Education, Culture, Sports, Science and Technology in Japan.

\bibliography{main}
\bibliographystyle{unsrt}

\end{document}